\documentclass{jfm}
\usepackage{graphicx}
\usepackage{newtxtext}
\usepackage{newtxmath}
\usepackage{natbib}
\usepackage{dashrule}
\usepackage{hyperref}
\hypersetup{
    colorlinks = true,
    urlcolor   = blue,
    citecolor  = black,
}

\usepackage{amstext}
\usepackage{amsmath}
\usepackage{amsfonts,amssymb}
\usepackage{framed}
\usepackage{algorithm}
\usepackage{algorithmicx}
\usepackage{algcompatible}
\usepackage{subcaption}
\usepackage{lipsum}
\usepackage{bm}
\usepackage{xcolor}
\usepackage{pdfpages}

\makeatletter
\DeclareRobustCommand{\sqcdot}{\mathbin{\mathpalette\morphic@sqcdot\relax}}
\newcommand{\morphic@sqcdot}[2]{%
  \sbox\z@{$\m@th#1\centerdot$}%
  \ht\z@=.2\ht\z@
  \vcenter{\box\z@}%
}
\makeatother

\def\L{\bm{\mbox{-----}}}
\def\dashL{\bm{\mbox{--~--~--}}}
\def\dashdotted{\xleaders\hbox to 1.0em{$- \sqcdot$}\hfill $-$}
\def\dotL{\bm{\mbox{$\sqcdot\ \hspace{-0.05 in} \sqcdot\ \hspace{-0.05 in} \sqcdot$}}}

\def\Lbox{{\bm{\mbox{---}}}~{\hspace*{-.1in} $\square$}\hspace*{-.1in}~{\bm{\mbox{---}}}}

\newcommand{\RomanNumeralCaps}[1]

\title{Phase-resolved ocean wave forecast with simultaneous current estimation through data assimilation}

\author{Guangyao Wang\aff{1}, Jinfeng Zhang\aff{1}, Yuxiang Ma\aff{2}, Qinghe Zhang\aff{1}, Zhilin Li\aff{1} \and Yulin Pan\aff{3}\corresp{\email{yulinpan@umich.edu}}}

\affiliation{\aff{1}State Key Laboratory of Hydraulic Engineering Simulation and Safety, Tianjin University, Tianjin 300072, China
\aff{2}State Key Laboratory of Coastal and Offshore Engineering, Dalian University of Technology, Dalian, 116023, China
\aff{3}Department of Naval Architecture and Marine Engineering, University of Michigan,
Ann Arbor, MI 48109, USA}

\begin{document}
\maketitle

\begin{abstract}
In Wang \& Pan (\emph{J. Fluid Mech., vol. 918, A19, 2021}), the authors developed the first ensemble-based data assimilation (DA) capability for the reconstruction and forecast of ocean surface waves, namely the EnKF-HOS method coupling an ensemble Kalman filter (EnKF) and the high-order spectral (HOS) method. In this work, we continue to enrich the method by allowing it to simultaneously estimate the ocean current field, which is in general not known \emph{a priori} and can (slowly) vary in both space and time. To achieve this goal, we incorporate the effect of ocean current (as unknown parameters) on waves to build the HOS-C method as the forward prediction model, and obtain a simultaneous estimation of (current) parameters and (wave) states via an iterative EnKF (IEnKF) method that is necessary to handle the complexity in this DA problem. The new algorithm, named IEnKF-HOS-C method, is first tested in synthetic problems with various forms (steady/unsteady, uniform/non-uniform) of current. It is shown that the IEnKF-HOS-C method is able to not only estimate the current field accurately, but also boost the prediction accuracy of the wave field (even) relative to the state-of-the-art EnKF-HOS method. Finally, using real data from a shipborne radar, we show that the IEnKF-HOS-C method successfully recovers the current speed that matches the \emph{in situ} measurement by a floating buoy. 
\end{abstract}

\begin{keywords}
Authors should not enter keywords on the manuscript, as these must be chosen by the author during the online submission process and will then be added during the typesetting process (see \href{https://www.cambridge.org/core/journals/journal-of-fluid-mechanics/information/list-of-keywords}{Keyword PDF} for the full list).  Other classifications will be added at the same time.
\end{keywords}

\section{Introduction}
In recent years, phase-resolved ocean wave models have received increasing attentions due to their close relevancy to the safety and efficiency of marine operations. Unlike traditional phase-averaged models \cite[e.g.][]{booij1999third,tolman2009user}, the phase-resolved models aim to predict individual waves, and therefore can capture the detailed information of the wave field (usually of $\mathcal{O}$(1$\text{km}^2$)) as a guidance for marine operations \cite[e.g.][]{ma2018wave,xiao2021time}. When nonlinear effects are considered, phase-resolved models have been  constructed via the high-order spectral (HOS) method \cite[][]{dommermuth1987high,west1987new}, including its later variants \cite[e.g.][]{craig1993numerical,xu2009numerical}, Zakharov equation \cite[][]{stuhlmeier2021deterministic} and machine learning techniques \cite[][]{mohaghegh2021rapid}. 

In spite of the prosperity of nonlinear wave models, their applications to wave forecast in realistic situations are limited due to the significant uncertainties that grow with time in forecasting the chaotic wave motion \cite[e.g.][]{janssen2008progress,annenkov2001predictability}. The source of the uncertainties include (i) the noisy initial conditions of the sea surface, which are usually taken from radar or buoy measurements with certain error characteristics; and (ii) the physical effects, say, of wind and ocean current that are not known \emph{a priori} and therefore not accurately accounted for in the nonlinear wave model. 

In addressing the issues of uncertainty growth, data assimilation (DA) methods have been developed (mostly in the field of geoscience as discussed in \cite{evensen2003ensemble,carrassi2018data}) which combines measurement data and model predictions to improve the analysis of the states. Among the available efforts of applying DA to phase-resolved wave forecast/analysis, most (if not all) focus mainly on addressing the uncertainty (i) from initial conditions or measurements, i.e., assuming the prediction model is perfect. These include methods based on variational DA \cite[][]{aragh2008variation,qi2018nonlinear,fujimoto2020ensemble,wu2022improved} that construct an initial condition to minimize the difference between predictions and measurements in future times, as well as methods based on the Kalman filter \cite[][]{yoon2015explicit} that solve for an optimal wave state at a particular time using prediction and data at the same time. Since the latter methods do not require future data for the analysis, they can be favorably applied in operational wave forecast (as a way to construct an optimal wave state, once data at the same time is available, that can be used as initial conditions for the forecast). Under this note, the first and last authors of this paper have developed the EnKF-HOS algorithm (as a substantial extension and improvement to \cite{yoon2015explicit}) which applies the ensemble Kalman filter coupled with ensemble HOS predictions for analysis and forecast of the ocean wave field. 

While the EnKF-HOS method (as the first ensemble-based DA method for phase-resolved ocean waves) has shown remarkable performances in extensive test cases, the uncertainty due to model parameters, i.e., the aforementioned uncertainty source (ii), is not considered except some very heuristic treatment through adaptive inflation (see \cite{wang2021phase} for details). This is a severe problem for ocean wave forecast as the wave evolution can be significantly affected by environmental parameters, e.g., the current and wind fields. One may think of determining these parameters from the global marine weather forecast, but it has to be realized that these global forecast results are usually only available at very coarse grid and sparse time instants. Therefore, a direct interpolation may result in significant errors and will certainly miss the the important spatial-temporal variation of these fields on the scales of the wave forecast domain and time horizon, e.g., rogue waves can be triggered as a wave train travels into an opposing current with an increasing current velocity \cite[e.g.][]{onorato2011triggering,ducrozet2021predicting}.

In this paper, we continue to develop the EnKF-HOS framework, enabling a simultaneous estimation of the wave states and model parameters. While the developed algorithm can in principle be applied to the estimation of different environmental parameters, we focus here on the ocean current field which can generally vary (slowly) in both space and time. To achieve this goal, we incorporate the current effect on waves to build the HOS-C method, following \cite{wang2018fully,pan2020model}, as the forward prediction model. When measurements of surface elevation are available, we then solve a DA problem that estimates both the (current) model parameters and (wave) states. We note that this is a non-trivial DA problem since the current parameters form a high-dimensional space (e.g., with the same dimensions as surface elevation in a most general setting) and can only be inferred from their correlation to the wave field (i.e., no direct measurement is available). Upon many trials we adopt an iterative ensemble Kalman filter (IEnKF) \cite[][]{iglesias2013ensemble,wang2016data}~which provides a satisfactory solution to this problem. The developed full method, named IEnKF-HOS-C, is first tested in a series of synthetic problems with various forms (steady/unsteady, uniform/non-uniform) of the current fields. It is shown that the IEnKF-HOS-C method not only provides an accurate estimation of the current field, but also boosts the wave analysis/forecast accuracy even compared to the state-of-the-art EnKF-HOS method. Finally, using real data of surface elevations from a shipborne radar, we show that the IEnKF-HOS-C method successfully recovers the current velocity that matches the \emph{in situ} measurement by a floating buoy.

The paper is organized as follows. The problem statement and detailed algorithm of IEnKF-HOS-C method are introduced in $\S$\ref{sec:PF}. The validation and benchmark of the method against synthetic cases and real marine radar data are presented in $\S$\ref{sec:numres}. We give a conclusion of the work in $\S$\ref{sec:conc}.

\section{Mathematical formulation and methodology}\label{sec:PF}

\subsection{Problem statement}
We consider the evolution of an ocean wave field under the effect of a surface current $\boldsymbol{U}(\boldsymbol{x},t)$, which in general can (slowly) vary in both the two-dimensional space $\boldsymbol{x}$ and time $t$. We have available a sequence of measurements of the ocean surface in spatial regions $\mathcal{M}_j$, with $j=0,1,2,3, \cdots$ the index of time $t$. In general, we allow $\mathcal{M}_j$ to be different for different $j$, reflecting a mobile system of measurement, e.g, a shipborne marine radar or moving probes. We denote the surface elevation and surface potential (for only waves), reconstructed from the measurements in $\mathcal{M}_j$, as $\eta_{\text{m},j}(\boldsymbol{x})$ and $\psi_{\text{m},j}(\boldsymbol{x})$, and assume that the error statistics associated with $\eta_{\text{m},j}(\boldsymbol{x})$ and $\psi_{\text{m},j}(\boldsymbol{x})$ is known \emph{a priori} from the inherent properties of the measurement equipment. 

In addition to the measurements, we have a yet-to-be-developed nonlinear wave model that is able to simulate the evolution of the surface waves (in particular surface elevation $\eta(\boldsymbol{x},t)$ and wave-only velocity potential $\psi(\boldsymbol{x},t)$) under the effect of $\boldsymbol{U}(\boldsymbol{x},t)$ given initial conditions. Our purpose is to incorporate measurements $\eta_{\text{m},j}(\boldsymbol{x})$ and $\psi_{\text{m},j}(\boldsymbol{x})$ into the model prediction sequentially (i.e., immediately as data become available in time) to simultaneously construct an optimized analysis of wave states $\big(\eta_{\text{a},j}(\boldsymbol{x}), \psi_{\text{a},j}(\boldsymbol{x})\big)$ and obtain an accurate inference/estimation of $\boldsymbol{U}(\boldsymbol{x},t)$. 

\subsection{The general IEnKF-HOS-C framework}
\label{sec:enkfhos}
Our new IEnKF-HOS-C method to solve the above problem is built upon the previous EnKF-HOS framework developed in \cite{wang2021phase}. In order to resolve the additional complexities associated with the current field $\boldsymbol{U}(\boldsymbol{x},t)$, the IEnKF-HOS-C method includes a number of new components (relative to the EnKF-HOS method): (i) a parameter-augmented state space $\big(\eta(\boldsymbol{x},t), \psi(\boldsymbol{x},t), \boldsymbol{U}(\boldsymbol{x},t)\big)$ which includes the current parameters; (ii) the HOS-C method which simulates the evolution of wave field under the effect of $\boldsymbol{U}(\boldsymbol{x},t)$, as well as a persistence model~\citep{notton2018forecasting,wu2019improving} $\partial\boldsymbol{U}(\boldsymbol{x},t)/\partial t=0$, used in the forecast step of the method; (iii) an iterative procedure in EnKF to build the IEnKF method which successfully handles the high-dimensional state/parameter estimation problem.
\begin{figure}
  \centerline{\includegraphics[scale =0.6]{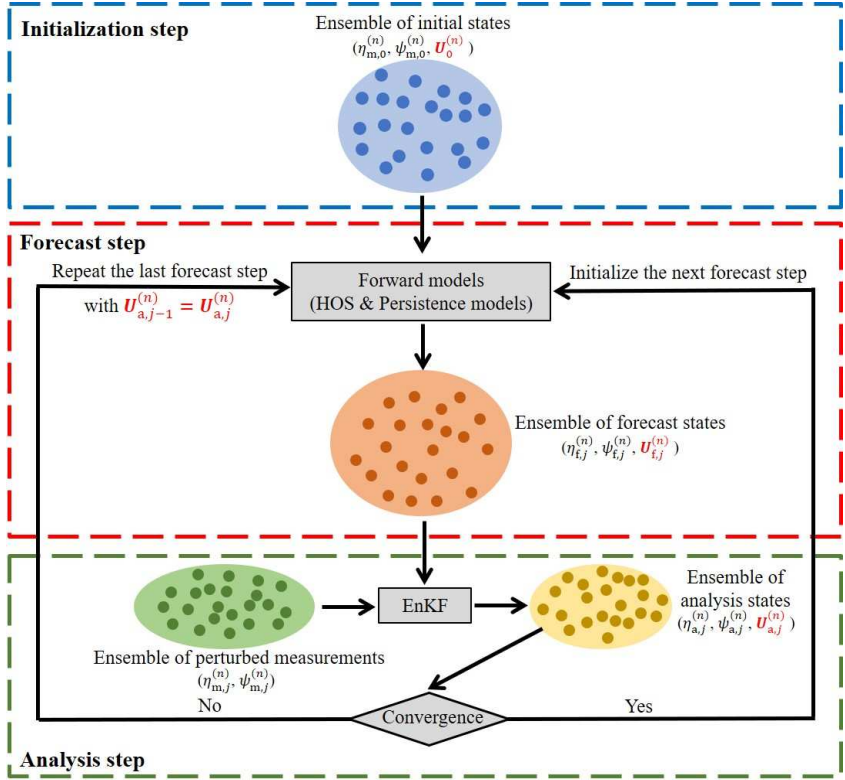}}
  \caption{Schematic illustration of the IEnKF-HOS-C coupled framework. The size of ellipse represents the amount of uncertainty.}
\label{fig:enkfhos}
\end{figure}
Figure \ref{fig:enkfhos} shows a schematic illustration of the new IEnKF-HOS-C framework. At initial time $t=t_0$, measurements $\eta_{\text{m},0}(\boldsymbol{x})$ and $\psi_{\text{m},0}(\boldsymbol{x})$ are available together with an initial guess $\boldsymbol{U}_0(\boldsymbol{x})$, based on which we generate ensembles of perturbed (augmented) states $\big(\eta_{\text{m},0}^{(n)}(\boldsymbol{x}), \psi_{\text{m},0}^{(n)}(\boldsymbol{x}), \boldsymbol{U}_0^{(n)}(\boldsymbol{x})\big)$, $n=1,2,...,N$, with $N$ the ensemble size. A forecast step is then performed, in which an ensemble of $N$ HOS-C and persistence-model simulations are conducted, taking $\big(\eta_{\text{m},0}^{(n)}(\boldsymbol{x}), \psi_{\text{m},0}^{(n)}(\boldsymbol{x}), \boldsymbol{U}_0^{(n)}(\boldsymbol{x})\big)$ as initial conditions for each ensemble member $n$, until $t=t_1$ when the next measurements become available. We note that the persistence model simply states that the forecast $\boldsymbol{U}_{\text{f},1}^{(n)}(\boldsymbol{x})=\boldsymbol{U}_0^{(n)}(\boldsymbol{x})$, but this is \emph{not} in contradiction with the inference of an unsteady current (see details in \S\ref{sec:hos}). At $t=t_1$, an analysis step is performed through EnKF where the model forecasts $\big(\eta_{\text{f},1}^{(n)}(\boldsymbol{x}), \psi_{\text{f},1}^{(n)}(\boldsymbol{x}),\boldsymbol{U}_{\text{f},1}^{(n)}(\boldsymbol{x})\big) $ are combined with new perturbed measurements $\big(\eta_{\text{m},1}^{(n)}(\boldsymbol{x}), \psi_{\text{m},1}^{(n)}(\boldsymbol{x})\big)$ to generate the analysis results $(\eta_{\text{a},1}^{(n)}(\boldsymbol{x}), \psi_{\text{a},1}^{(n)}(\boldsymbol{x}), \boldsymbol{U}_{\text{a},1}^{(n)}(\boldsymbol{x}))$. Since $\boldsymbol{U}_{\text{a},1}^{(n)}$ is obtained only through its correlation to the wave field (i.e., no direct measurement) and the forecast step has been possibly performed with an inaccurate current field (i.e., $\boldsymbol{U}_0(\boldsymbol{x})\neq \boldsymbol{U}^{\text{true}}(\boldsymbol{x},t_0)$), it is necessary to conduct iterations between the forecast and analysis steps to facilitate the convergence of the analyzed current field to the true situation. In particular, we iterate the forecast and analysis steps, every time using the new estimation $\boldsymbol{U}_{\text{a},1}^{(n)}(\boldsymbol{x})$ in analysis to replace $\boldsymbol{U}_0^{(n)}(\boldsymbol{x})$ in forecast until a desired tolerance is reached.

With $\big(\eta_{\text{a},1}^{(n)}(\boldsymbol{x}), \psi_{\text{a},1}^{(n)}(\boldsymbol{x}), \boldsymbol{U}_{\text{a},1}^{(n)}(\boldsymbol{x})\big)$ available after IEnKF, they are taken as initial conditions for a new ensemble of HOS-C and persistence-model simulations, and the procedures are repeated for $t=t_2, t_3, \cdots$ until the desired operation time $t_{\text{max}}$ is reached. We next describe in detail the key components in the IEnKF-HOS-C method, including the generation of measurement ensembles and state augmentation (\S\ref{sec:pert}), the HOS-C method and persistence model (\S\ref{sec:hos}), and the IEnKF procedure (\S\ref{sec:enkf}). For simplicity, in the following description we assume that the gravitational acceleration and fluid density are unity (so that they do not appear in equations) by choices of proper time and mass units.

\subsection{Generation of measurement ensembles and state augmentation}
\label{sec:pert}
As described in \S\ref{sec:enkfhos}, ensembles of perturbed measurements of the surface elevation $\left\{ \eta^{(n)}_{\text{m},j}(\boldsymbol{x})\right\}_{n=1}^N$ and velocity potential $\left\{\psi^{(n)}_{\text{m},j}(\boldsymbol{x})\right\}_{n=1}^N$ are needed at both the initialization ($j=0$) and analysis ($j=1,~2,~3\cdots$) steps. In addition, an ensemble of initial current velocity $\left\{ \boldsymbol{U}^{(n)}_{0}(\boldsymbol{x})\right\}_{n=1}^N$ is needed (at $t=t_0$) as a state augmentation to start the full IEnKF-HOS-C algorithm.

To illustrate the generation of these ensembles, it is convenient to first define a random field $w(\boldsymbol{x})$ as a zero-mean Gaussian process with spatial correlation function~\citep{evensen2003ensemble,evensen2009data} 
\begin{equation}
C(w(\boldsymbol{x}_1),w(\boldsymbol{x}_2))=\begin{cases}c_w\exp\left(-\displaystyle\frac{|\boldsymbol{x}_1-\boldsymbol{x}_2|^2}{a_w^2}\right) &\text{for}~ |\boldsymbol{x}_1-\boldsymbol{x}_2|\leq\sqrt{3}a_w,\\
0~&\text{for}~|\boldsymbol{x}_1-\boldsymbol{x}_2|>\sqrt{3}a_w.
\end{cases}
\label{eq:noise1}
\end{equation} 
with $c_w$ the variance of $w(\boldsymbol{x})$ and $a_w$ the de-correlation length scale.

Following the method proposed by~\cite{wang2021phase}, we first produce $\eta_{\text{m},j}^{(n)}$ by adding a random-field perturbation $w(\boldsymbol{x})$ (defined by \eqref{eq:noise1} and with different realizations for different $n$) to $\eta_{m,j}$, i.e.,
\begin{equation}
\eta_{\text{m},j}^{(n)}(\boldsymbol{x})={\eta}_{\text{m},j}(\boldsymbol{x})+w(\boldsymbol{x}),
\label{eq:etaini}
\end{equation}
and construct the surface potential $\psi_{\text{m},j}^{(n)}$ by linear wave theory
\begin{equation}
 \psi_{\text{m},j}^{(n)}(\boldsymbol{x})\sim \int\frac{i}{ \sqrt{\mid\boldsymbol{k}\mid}}\tilde{\eta}_{\text{m},j}^{(n)}(\boldsymbol{k})e^{i\boldsymbol{k}\cdot\boldsymbol{x}}d\boldsymbol{k},
 \label{eq:lpsi}
\end{equation}
where $\tilde{\eta}^{(n)}_{\text{m},j}(\boldsymbol{k})$ denotes the Fourier coefficient of the $n_{th}$ member of the perturbed surface elevation at vector wavenumber $\boldsymbol{k}$. We note that \eqref{eq:lpsi} is a direct result of the linear wave equation, and it is not modified by the presence of a uniform current (since physically the current does not affect the velocity field of waves except for a Doppler shift).

To generate the ensemble $\left\{ \boldsymbol{U}^{(n)}_{0}(\boldsymbol{x})\right\}_{n=1}^N\equiv \left\{ U^{(n)}_{x,0}(\boldsymbol{x}), U^{(n)}_{y,0}(\boldsymbol{x}) \right\}_{n=1}^N$, we start from an initial guess $\boldsymbol{U}_0(\boldsymbol{x}) \equiv \big(U_{x,0}(\boldsymbol{x}),U_{y,0}(\boldsymbol{x})\big)$ which is in general not the same as the truth $\boldsymbol{U}^{\text{true}}(\boldsymbol{x},t_0)$. In practice, $\boldsymbol{U}_0(\boldsymbol{x})$ can be set as zero or taken from the results of large-scale marine weather forecast. We generate the ensemble of current field by adding another random-field perturbation $u(\boldsymbol{x})$ to each component of  $\boldsymbol{U}_0(\boldsymbol{x})$, i.e., 
\begin{equation}
    U_{*,0}^{(n)}(\boldsymbol{x})= U_{*,0}(\boldsymbol{x})+u(\boldsymbol{x}),
\label{eq:ustar}
\end{equation}
where the subscript $*$ represents $x$ or $y$, and $u(\boldsymbol{x})$ is a random field defined by \eqref{eq:noise1} with $u$ replacing $w$, i.e., with variance $c_u$ and de-correlation length scale $a_u$.

\subsection{HOS-C method and persistence model}
\label{sec:hos}
Given the initial conditions $\big(\eta_{\text{m},0}^{(n)}(\boldsymbol{x}), \psi_{\text{m},0}^{(n)}(\boldsymbol{x}), \boldsymbol{U}_0^{(n)}(\boldsymbol{x})\big)$ or $\big(\eta_{\text{a},j}^{(n)}(\boldsymbol{x}), \psi_{\text{a},j}^{(n)}(\boldsymbol{x}), \boldsymbol{U}_{\text{a},j}^{(n)}(\boldsymbol{x})\big)$ with $j \geq 1$, for each ensemble member $n$, the evolution of the (wave and current) augmented state from $t_j$ to $t_{j+1}$ is solved by integrating a nonlinear wave equation under the effect of the current:
\begin{eqnarray}
    \frac{\partial\eta(\boldsymbol{x},t)}{\partial t} &+& \frac{\partial\psi(\boldsymbol{x},t)}{\partial \boldsymbol{x}} \cdot \frac{\partial\eta(\boldsymbol{x},t)}{\partial \boldsymbol{x}} -\left[1+\frac{\partial\eta(\boldsymbol{x},t)}{\partial \boldsymbol{x}}\cdot \frac{\partial\eta(\boldsymbol{x},t)}{\partial \boldsymbol{x}}\right]\phi_z(\boldsymbol{x},t)\nonumber\\ 
    &+&\frac{\partial\eta(\boldsymbol{x},t)}{\partial \boldsymbol{x}}\cdot {\boldsymbol{U}}(\boldsymbol{x},t_j)+\eta(\boldsymbol{x},t)\frac{\partial}{\partial \boldsymbol{x}}\cdot {\boldsymbol{U}}(\boldsymbol{x},t_j)=0,
\label{eq:bc1}
\end{eqnarray}

\begin{eqnarray}
&&\frac{\partial\psi(\boldsymbol{x},t)}{\partial t} + \frac{1}{2}\frac{\partial\psi(\boldsymbol{x},t)}{\partial \boldsymbol{x}}\cdot\frac{\partial\psi(\boldsymbol{x},t)}{\partial \boldsymbol{x}}+\eta(\boldsymbol{x},t)\nonumber\\ 
&&-\frac{1}{2}\left[1+\frac{\partial\eta(\boldsymbol{x},t)}{\partial \boldsymbol{x}}\cdot \frac{\partial\eta(\boldsymbol{x},t)}{\partial \boldsymbol{x}}\right]\phi_z(\boldsymbol{x},t)^2+\frac{\partial\psi(\boldsymbol{x},t)}{\partial \boldsymbol{x}}\cdot {\boldsymbol{U}}(\boldsymbol{x},t_j)=0,
\label{eq:bc2}
\end{eqnarray}
and a persistence model
\begin{equation}
    \frac{\partial\boldsymbol{U}(\boldsymbol{x},t)}{\partial t}=0.
    \label{eq:persis}
\end{equation}
In \eqref{eq:bc1} and \eqref{eq:bc2}, $\phi_z(\boldsymbol{x},t)\equiv \partial \phi/\partial z|_{z=\eta}(\boldsymbol{x},t)$ is the surface vertical velocity with $\phi(\boldsymbol{x},z,t)$ being the velocity potential of the wave field, and $\psi(\boldsymbol{x},t)\equiv\phi(\boldsymbol{x},\eta,t)$. The variable ${\boldsymbol{U}}(\boldsymbol{x},t_j)$ in the equations should be considered as the estimated quantity, taking either $\boldsymbol{U}_0^{(n)}(\boldsymbol{x})$ at $t=t_0$ or $\boldsymbol{U}_{\text{a},j}^{(n)}(\boldsymbol{x})$ at $t=t_j$.
The two equations \eqref{eq:bc1} and \eqref{eq:bc2} describe the evolution of a nonlinear wave field under the effect of an irrotational current which slowly varies in space \cite[][]{wang2018fully,pan2020model}. As discussed in \cite{pan2020model}, this set of equations form a Hamiltonian system conserving the total energy of the wave and current, and it is possible to relax the scale-separation assumption and irrotational assumption with more developments at certain situations. 

The persistence model \eqref{eq:persis} simply states that the current field remains steady in the forecast step, i.e., $\boldsymbol{U}_{\text{f},j+1}=\boldsymbol{U}_{\text{a},j}$ (see applications in other contexts, e.g.,~\cite{santitissadeekorn2015two,notton2018forecasting,wu2019improving}). We remark that this is \emph{not} in contradiction with the estimation of an unsteady current field which slowly varies in time but can be approximated as a constant in the forecast interval (from $t_j$ to $t_{j+1}$). In fact, the time variation of the unsteady current is captured in the IEnKF procedure that will be discussed in the next section.   

\subsection{Data Assimilation Scheme by IEnKF}
\label{sec:enkf}
Let's now assume that we have obtained the forecast ensemble $\left\{ \eta^{(n)}_{\text{f},j}\right\}_{n=1}^N$, $\left\{ \psi^{(n)}_{\text{f},j}\right\}_{n=1}^N$, and $\left\{ \boldsymbol{U}^{(n)}_{\text{f},j}\right\}_{n=1}^N$ by integrating \eqref{eq:bc1}$\sim$\eqref{eq:persis} from $t_{j-1}$ to $t_j$. To describe the analysis step, we first introduce the notation of a covariance operator
\begin{equation}
    \mathfrak{C}(x,y)=\frac{1}{N-1}\sum_{n=1}^N(x^{(n)}-\bar x)(y^{(n)}-\bar y)^T
    \label{eq:cov}
\end{equation}
which produces the covariance matrix between two vectors $x$ and $y$ through ensemble average, with the overbar in the equation denoting the ensemble mean.

The analysis step combines $\left(\eta^{(n)}_{\text{f},j}\in \mathbb{R}^L,\psi^{(n)}_{\text{f},j}\in \mathbb{R}^L,\boldsymbol{U}^{(n)}_{\text{f},j}\in \mathbb{R}^{2L}\right)$ and $\left(\eta^{(n)}_{\text{m},j}\in \mathbb{R}^d,\psi^{(n)}_{\text{m},j}\in \mathbb{R}^d\right)$ with $L$ and $d$ being the dimensions of model (forecast) space and measurement space respectively. Through EnKF (no iteration yet), this step can be formulated as
\begin{equation}
{\eta^{(n)}_{\text{a},j}}={\eta^{(n)}_{\text{f},j}}+{\boldsymbol{Q}_{\eta\eta,j}}{\boldsymbol{G}^\text{T}}\left(\boldsymbol{G}\boldsymbol{Q}_{\eta\eta,j}\boldsymbol{G}^\text{T}+\boldsymbol{R}_{\eta\eta,j}\right)^{-1} \left({\eta^{(n)}_{\text{m},j}}-{\boldsymbol{G}} {\eta^{(n)}_{\text{f},j}}\right),
\label{eq:ana1}
\end{equation}

\begin{equation}
{\psi^{(n)}_{\text{a},j}}={\psi^{(n)}_{\text{f},j}}+{\boldsymbol{Q}_{\psi\psi,j}}{\boldsymbol{G}^\text{T}}\left(\boldsymbol{G}\boldsymbol{Q}_{\psi\psi,j}\boldsymbol{G}^\text{T}+\boldsymbol{R}_{\psi\psi,j}\right)^{-1} \left({\psi^{(n)}_{\text{m},j}}-{\boldsymbol{G}} {\psi^{(n)}_{\text{f},j}}\right),
\label{eq:ana2}
\end{equation}

\begin{equation}
{\boldsymbol{U}^{(n)}_{\text{a},j}}={\boldsymbol{U}^{(n)}_{\text{f},j}}+{\boldsymbol{Q}_{U\eta,j}}{\boldsymbol{G}^\text{T}}\left(\boldsymbol{G}\boldsymbol{Q}_{\eta\eta,j}\boldsymbol{G}^\text{T}+\boldsymbol{R}_{\eta\eta,j}\right)^{-1} \left({\eta^{(n)}_{\text{m},j}}-{\boldsymbol{G}} {\eta^{(n)}_{\text{f},j}}\right),
\label{eq:ana3}
\end{equation}
where 
\begin{equation}
{\boldsymbol{Q}_{xy,j}}=\mathfrak{C}(x_{\text{f},j}, y_{\text{f},j}),
\label{eq:Q}
\end{equation}
\begin{equation}
{\boldsymbol{R}_{xy,j}}=\mathfrak{C}(x_{\text{m},j}, y_{\text{m},j}).
\label{eq:R}
\end{equation}
The operator (or matrix) $\boldsymbol{G}:\mathbb{R}^L \rightarrow \mathbb{R}^d$ is an observation operator mapping the $L$-dimensional model space to the $d$-dimensional measurement space, which is constructed by a linear interpolation in this study (e.g., for measurements on grid points, $\boldsymbol{G}$ is reduced to an operation to take the corresponding elements in the model vector).
\begin{algorithm*}[htpb!]
\centering
\caption{Algorithm for IEnKF-HOS-C method}

\label{al:asimilation}
\begin{algorithmic}[1]
\State {\bf{Input}}: ${\eta_{\text{m},0}}$, ${\psi_{\text{m},0}}$, $\boldsymbol{U}_0$, $t_{\text{max}}$, $N$, $\delta$, $h_{\text{max}}$
\State {\bf{Begin}}
\State initialize:\\
\hspace{0.6cm}       $t=t_0, j=0$\\
\hspace{0.6cm}       Generate $\eta^{(n)}_{\text{m},0}(\boldsymbol{x}),~\psi^{(n)}_{\text{m},0}(\boldsymbol{x})$, and $\boldsymbol{U}^{(n)}_{0}(\boldsymbol{x})$ with \eqref{eq:noise1}$\sim$\eqref{eq:ustar}
\State time loop:
\State \hspace{0.6cm} {\bf{while}} $t \leq t_{\text{max}} $ {\bf{do}}\
\State \hspace{1.0cm} $j=j+1$, $h=1$\
\State \hspace{1.0 cm} {\bf{read}} $\eta_{\text{m},j}$
\State \hspace{1.0cm} Generate $ \eta^{(n)}_{\text{m},j}(\boldsymbol{x})$ and~$\psi^{(n)}_{\text{m},j}(\boldsymbol{x})$ with \eqref{eq:noise1}$\sim$\eqref{eq:lpsi}\
\State \hspace{1.0cm} {\bf{while}} $h \leq h_{\text{max}} $ {\bf{do}}\
\State \hspace{1.4cm} $h=h+1$
\State \hspace{1.4cm} Solve \eqref{eq:bc1}$\sim$\eqref{eq:persis} until $t=t_j$ to obtain $\eta^{(n)}_{\text{f},j}(\boldsymbol{x})$,~$\psi^{(n)}_{\text{f},j}(\boldsymbol{x})$, and~$\boldsymbol{U}^{(n)}_{\text{f},j}(\boldsymbol{x})$
\State \hspace{1.4cm} Calculate $\eta^{(n)}_{\text{a},j}(\boldsymbol{x})$,~$\psi^{(n)}_{\text{a},j}(\boldsymbol{x})$, and $\boldsymbol{U}^{(n)}_{\text{a},j}(\boldsymbol{x})$ with~\eqref{eq:ana1}$\sim$\eqref{eq:ana3}
\State \hspace{1.4cm} {\bf{if}}~$||\bar{\boldsymbol{U}}_{\text{f},j}(\boldsymbol{x})-\bar{\boldsymbol{U}}_{\text{a},j}(\boldsymbol{x})||_2 < \delta$~{\bf{then}}

\State \hspace{1.8cm} {\bf{break}}
\State \hspace{1.4cm} {\bf{else}}
\State \hspace{1.8cm} $\boldsymbol{U}^{(n)}_{\text{a},j-1}(\boldsymbol{x})=\boldsymbol{U}^{(n)}_{\text{a},j}(\boldsymbol{x})$
\State \hspace{1.4cm} {\bf{endif}}
\State \hspace{1.0cm} {\bf{end}}
\State \hspace{1.0cm} {\bf{Output}} $\bar{\eta}_{\text{a},j}(\boldsymbol{x})$,~$\bar{\psi}_{\text{a},j}(\boldsymbol{x})$, and~ $\bar{\boldsymbol{U}}_{\text{a},j}(\boldsymbol{x})$
\State \hspace{0.6cm} {\bf{end}}
\State {\bf{end}}
\end{algorithmic}
\end{algorithm*}
We note that \eqref{eq:ana1} and \eqref{eq:ana2} are equivalent to the analysis equation (2.13) in \cite{wang2021phase} written in a form of ensemble matrix. Equation \eqref{eq:ana3} provides the update (thus estimation) of the current field, which is achieved through its correlation to the surface elevation field (i.e., the matrix $\boldsymbol{Q}_{U\eta,j}$ established through ensembles of HOS-C forecast). In addition, \eqref{eq:ana1} and \eqref{eq:ana3} combined are equivalent to a standard EnKF equation for an augmented state vector of $(\eta,\boldsymbol{U})$. On the other hand, while it is also possible to estimate $\boldsymbol{U}$ through its correlation with $\psi$, this alternative approach is not more beneficial to \eqref{eq:ana3} from both first-principle reasoning and our numerical tests.

Let us next consider the situation that $\boldsymbol{U}_{\text{f},j}(\boldsymbol{x})$ is different from the true field $\boldsymbol{U}^{\text{true}}(\boldsymbol{x},t_j)$. While the analysis ${\boldsymbol{U}}^{(n)}_{\text{a},j}(\boldsymbol{x})$ may provide an update that is closer to the truth, the previous forecast step from $j-1$ to $j$ has been performed with an inaccurate current field. To remedy this situation, it is necessary to perform iterations between the forecast and analysis steps. In particular, for each iteration we replace $\boldsymbol{U}^{(n)}_{\text{a},j-1}(\boldsymbol{x})$ by $\boldsymbol{U}^{(n)}_{\text{a},j}(\boldsymbol{x})$ and repeat the forecast (with updated current field) and analysis steps, until convergence is achieved with a criterion
\begin{equation}
    ||\bar{\boldsymbol{U}}_{\text{f},j}(\boldsymbol{x})-\bar{\boldsymbol{U}}_{\text{a},j}(\boldsymbol{x})||_2 < \delta,
\label{eq:tol}
\end{equation}
or if a preset maximum number of iterations $h_{\text{max}}$ is reached.
We have now completed the description of the IEnKF-HOS-C method, with a pseudo-code provided in Algorithm~\ref{al:asimilation}. In addition, in implementation of IEnKF-HOS-C other practical procedures are required, including the adaptive inflation and localization schemes, and the treatment of the mismatch between the predictable and measurement regions. These procedures are discussed in detail in \cite{wang2021phase} and will not be re-presented in this paper.

\section{Results}
\label{sec:numres}
We validate the IEnKF-HOS-C method through a series of test cases with both synthetic and real ocean wave fields. For each case of the former, a reference HOS-C simulation is conducted with a prescribed current field to produce the true wave solution, onto which random errors are superposed to generate the synthetic noisy measurements. For the latter, we make use of the real wave data obtained from an onboard Doppler coherent marine radar~\citep{lyzenga2015real,nwogu2010surface}, with the reference current velocity measured by a floating buoy. For both types of cases, we use $N=100$ ensemble members in the IEnKF-HOS-C method.

The performance of the IEnKF-HOS-C method can be evaluated by a natural metric of the estimated current field, which should be compared to the reference solutions (prescribed true current fields in the synthetic cases and buoy measurement in the real case). In addition, for the synthetic cases, since the true wave solution is known, another metric can be defined as the error of the analyzed wave field relative to the true solution:
\begin{equation}
\epsilon(t)=\frac{\int_\mathcal{A}\mid\eta^\text{true}(\boldsymbol{x},t)-\eta^{\text{sim}}(\boldsymbol{x},t)\mid^2 d\mathcal{A}}{2\sigma_{\eta}^2\mathcal{A}},
\label{eq:epsilon}
\end{equation}
where $\mathcal{A}$ is the area of the simulation region, $\sigma^2_\eta$ is the variance of the reference surface elevation field, $\eta^{\text{true}}(\boldsymbol{x},t)$ and $\eta^{\text{sim}}(\boldsymbol{x},t)$ represent respectively the true (reference) surface elevation field and the simulation results (that will be obtained through IEnKF-HOS-C, HOS-C-only and our previous EnKF-HOS methods for comparison).
 
\subsection{Synthetic cases}
We consider synthetic cases of both two-dimensional (2D, with one horizontal direction $x$) and three-dimensional (3D, with two horizontal directions $\boldsymbol{x}=(x,y)$) wave fields. The true wave solution $\eta^{\text{true}}(\boldsymbol{x},t)$ for each case is generated by a reference HOS-C simulation with an exact initial condition and a prescribed current field. For the 2D case, we use a reference initial wave field described by a JONSWAP spectrum with a peak wavenumber $k_p=16k_0$ (with $k_0$ the fundamental wavenumber), a global steepness $k_pH_s/2=0.11$ (with $H_s$ the significant wave height) and an enhancement factor $\gamma=3.3$. For the 3D case, the initial wave field is taken from the same spectrum together with a directional spreading function
\begin{equation}
    D(\theta)=
    \begin{cases}
    \displaystyle\frac{2}{\beta}\cos^2(\displaystyle\frac{\pi}{\beta}\theta),~&\text{for}~\displaystyle-\frac{\beta}{2}<\theta<\displaystyle\frac{\beta}{2} \\
    0,~&\text{otherwise}
    \end{cases}
\end{equation}
where $\beta=\pi/6$ is the spreading angle. Without loss of generality, we assume that the true current velocity is always along the $x$-direction, expressed as 
\begin{equation}
    \boldsymbol{U}^{\text{true}}(\boldsymbol{x})=(U_x^{\text{true}}(\boldsymbol{x}), U_y^{\text{true}}(\boldsymbol{x})), 
\end{equation}
with $U_y^{\text{true}}(\boldsymbol{x})=0$ which needs to be estimated together with the non-zero component $U_x^{\text{true}}(\boldsymbol{x})$ in 3D cases through IEnKF-HOS-C. We remark that in making $U_y^{\text{true}}(\boldsymbol{x})=0$ it is assumed that the incompressibility of the current field, if required, is satisfied through the balance of gradients between $x$-direction and vertical motions. This assumption brings conveniences in validating the estimated velocity field, and does not considerably deteriorate the generality of the validation.  

To generate the noisy measurements of the wave field, we first superpose a random field onto the reference solution of surface elevation
\begin{equation}
    \eta_{\text{m}}(\boldsymbol{x})=\eta^\text{true}(\boldsymbol{x})+w(\boldsymbol{x}),
    \label{eq:mea}
\end{equation}
where $w(\boldsymbol{x})$ is sampled with $c_w=0.0025\sigma^2_{\eta}$ and $a_w=\pi/2$ in~\eqref{eq:noise1}. $\psi_{\text{m}}(\boldsymbol{x})$ is then generated based on the linear wave theory (similar to the generation of $\psi^{(n)}_{\text{m}}(\boldsymbol{x})$ from $\eta^{(n)}_{\text{m}}(\boldsymbol{x})$ shown in \eqref{eq:lpsi})
\begin{equation}
\psi_{\text{m}}(\boldsymbol{x})\sim\int\frac{i}{ \sqrt{\mid\boldsymbol{k}\mid}}\tilde{\eta}_{\text{m}}(\boldsymbol{k})e^{i\boldsymbol{k}\cdot\boldsymbol{x}}d\boldsymbol{k},
\label{eq:e2}
\end{equation}
where $\tilde{\eta}_{\text{m}}(\boldsymbol{k})$ denotes the measured surface elevation in the Fourier space. Regarding the initial guess of current velocity ${U}_{*,0}$, we assume it to be uniform, and reasonable to some extent (with its deviation from $U_*^{\text{true}}$ not too large) due to the existence of marine weather forecast information in practice. 

Finally, on the computation side, we use $L=256$ grid points with a spatial domain $\mathcal{A}=[0,2\pi)$ for 2D simulations, and $L=64\times64$ grid points with $\mathcal{A}=[0,2\pi)\times[0,2\pi)$ for 3D simulations, respectively. We will next describe all synthetic cases classified by the form of the current fields that can be uniform/non-uniform and steady/unsteady.
 
\subsubsection{Results for uniform current fields}
We start from relatively simple cases with a uniform current field that can be steady or unsteady. Under this situation, the current field to be estimated at each time instant is reduced to a scalar for the 2D cases or a vector with two components for the 3D cases. In the analysis step, the equation \eqref{eq:ana3} can be modified accordingly in dimensions for the simplified scalar (vector) velocity fields. 

\vspace{0.5cm}
\begin{description}
   \item[Steady cases]
\end{description}
We consider both 2D and 3D wave fields with the true current velocity given by 
\begin{equation}
   U_x^{\text{true}}=0.25v_p(k_p),
\end{equation}
where $v_p(k_p)$ represents the phase velocity of the peak wavenumber $k_p$ in the JONSWAP spectrum. The initial guess of the current velocity is predefined as 
\begin{equation}
    {U}_{x,0}=0.3v_p(k_p)
    \label{eq:ux}
\end{equation}
for the 2D case, and
\begin{equation}
     \boldsymbol{U}_{0}=\big(0.3v_p(k_p),~ 0.05v_p(k_p)\big) 
    \label{eq:uy}
\end{equation}
for the 3D case. We note that a non-zero $y$-component is used for $\boldsymbol{U}_{0}$ even though the truth is zero. The ensemble of the current velocity is generated by superposing zero-mean random errors, which are sampled from \eqref{eq:noise1} with $c_u=0.04(U_x^{\text{true}})^2$ and $a_u \rightarrow \infty$, onto each component of the initial guess.

For the IEnKF-HOS-C simulations, data from $d=12$ locations (randomly selected with a uniform distribution) are assimilated with an interval $\tau=T_p/32$, with $T_p$ the wave period for the peak mode $k_p$. The simulations start from the noisy measurements of the initial wave field (which are generated by \eqref{eq:mea} and \eqref{eq:e2}) and the initial guess of the current velocity, and are conducted until $t=50T_p$.

\begin{figure}
     \centering
     \begin{subfigure}[b]{1\textwidth}
         \centering
         \includegraphics[trim=0cm 0cm 0cm 0cm, clip,scale=0.3]{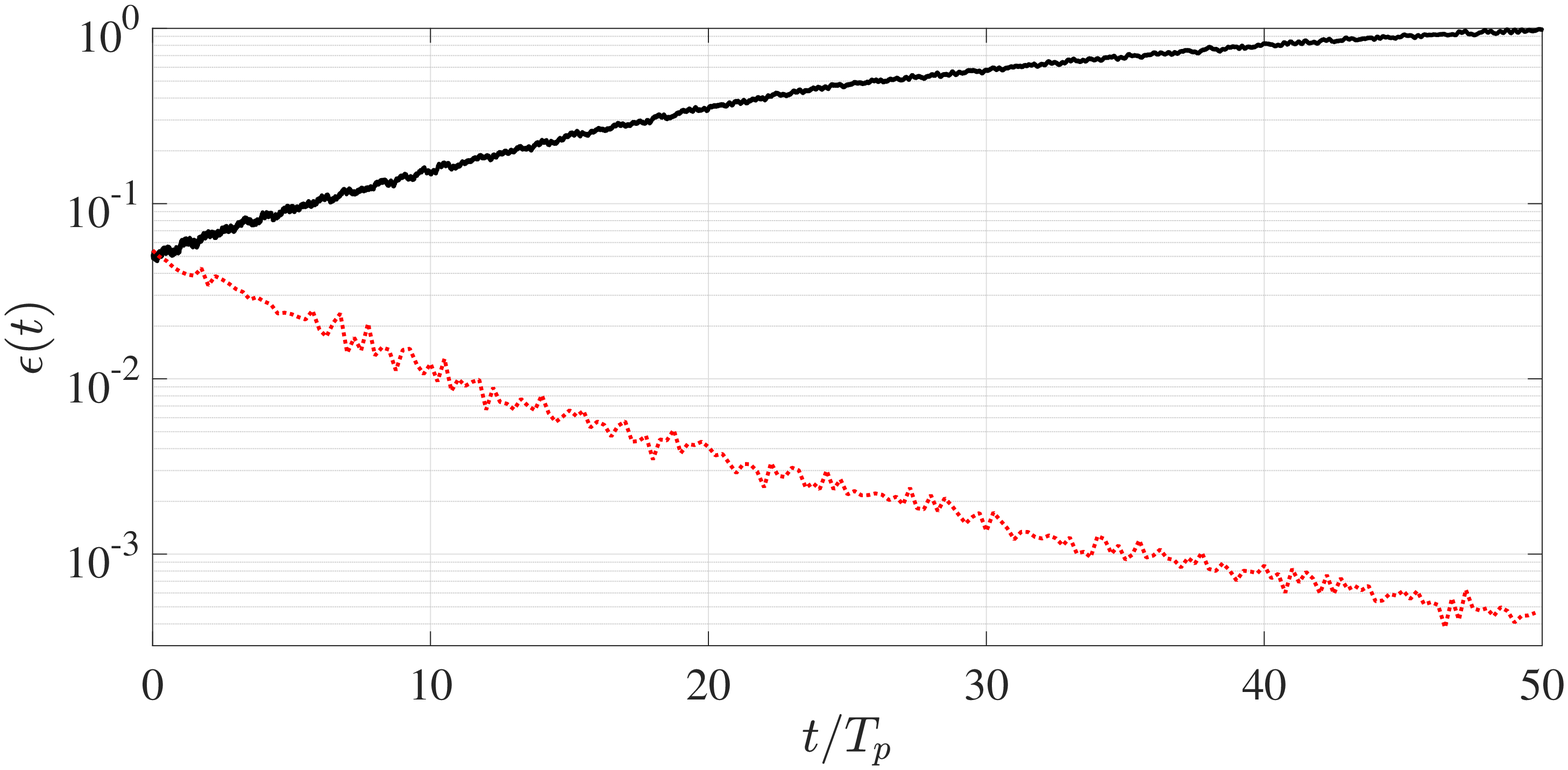}
         \caption{}
         \label{fig:e2d}
     \end{subfigure}
     \begin{subfigure}[b]{1\textwidth}
         \centering
         \includegraphics[trim=0cm 0cm 0cm 0cm, clip,scale=0.3]{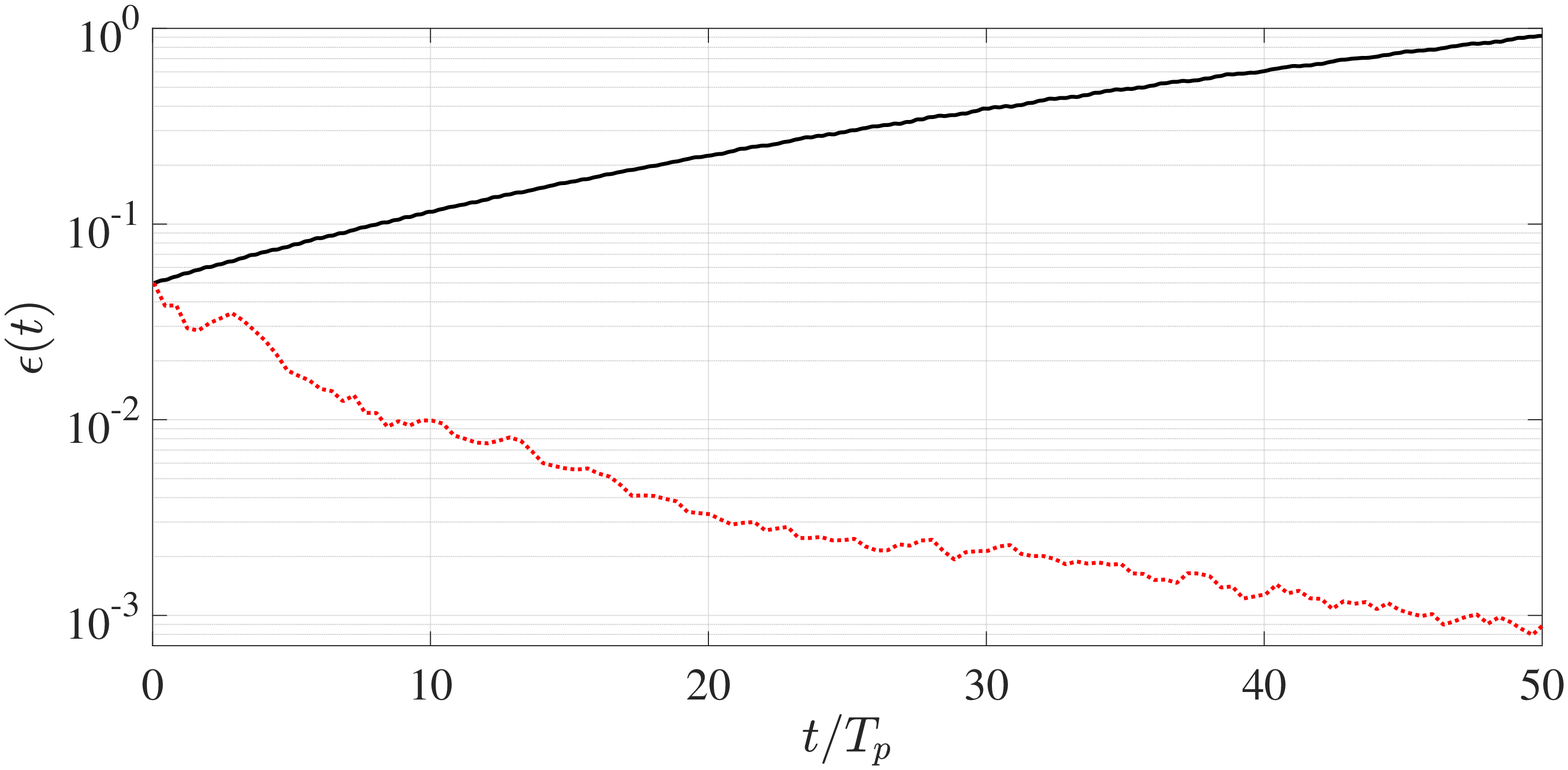}
         \caption{}
         \label{fig:e3d}
     \end{subfigure}
        \caption{Errors $\epsilon(t)$ from IEnKF-HOS-C ({\color{red}\dotL}) and HOS-C-only (\rule[0.5ex]{0.5cm}{0.5pt}) methods, for (a) 2D and (b) 3D cases with a steady and uniform current field.}
        \label{fig:e2d3d}
\end{figure}

The errors $\epsilon(t)$ obtained from IEnKF-HOS-C and HOS-C-only simulations (starting from the same wave field and initial guess of current velocity) are shown in figure~\ref{fig:e2d3d}. For both 2D and 3D cases, as the simulation proceeds $\epsilon(t)$ obtained from the HOS-C-only method increases from the initial value $\epsilon(0)\approx 0.05$ and approaches $\mathcal{O}(1)$ at $t/T_p\approx 50$; whereas with the IEnKF-HOS-C method, $\epsilon(t)$ keeps decreasing as the measurements are assimilated sequentially, and becomes two orders of magnitude smaller than its initial value at the end of simulations. 

\begin{figure}
     \centering
     \begin{subfigure}[b]{1\textwidth}
         \centering
         \includegraphics[trim=0cm 0cm 0cm 0cm, clip,scale=0.3]{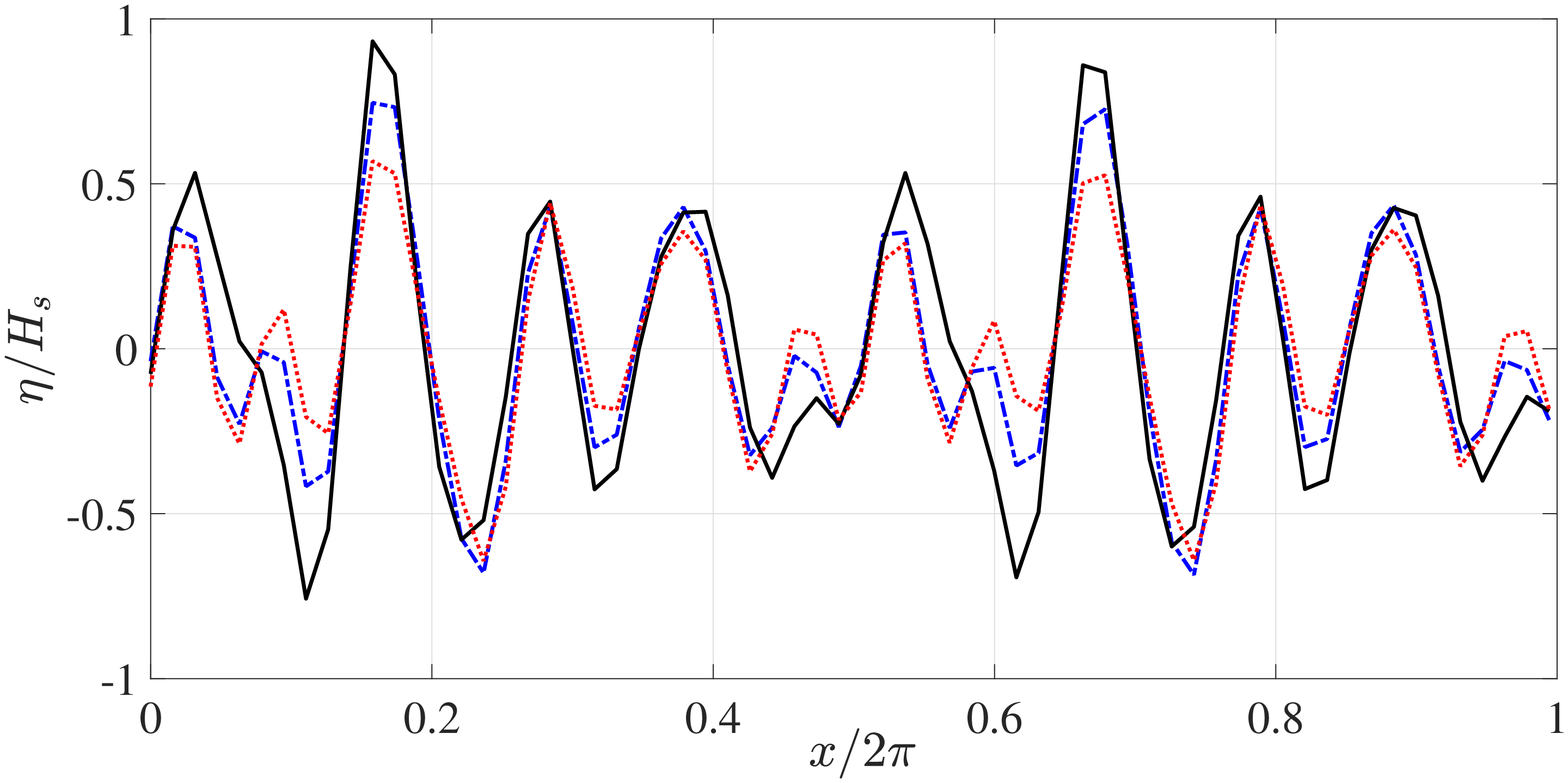}
         \caption{}
         \label{fig:wave2d_5}
     \end{subfigure}
     \begin{subfigure}[b]{1\textwidth}
         \centering
         \includegraphics[trim=0cm 0cm 0cm 0cm, clip,scale=0.3]{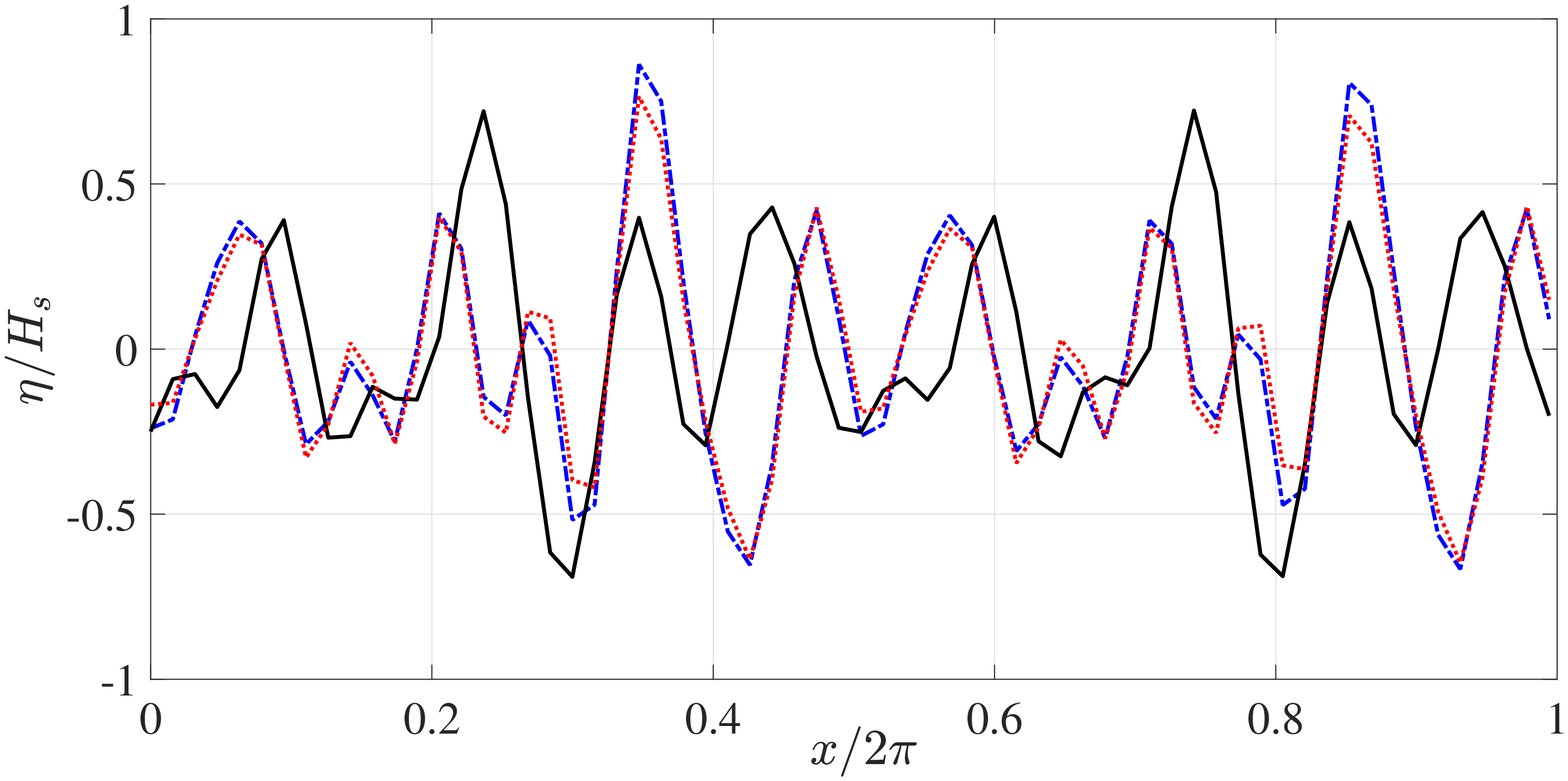}
         \caption{}
         \label{fig:wave2d_25}
     \end{subfigure}
     
     \begin{subfigure}[b]{1\textwidth}
         \centering
         \includegraphics[trim=0cm 0cm 0cm 0cm, clip,scale=0.3]{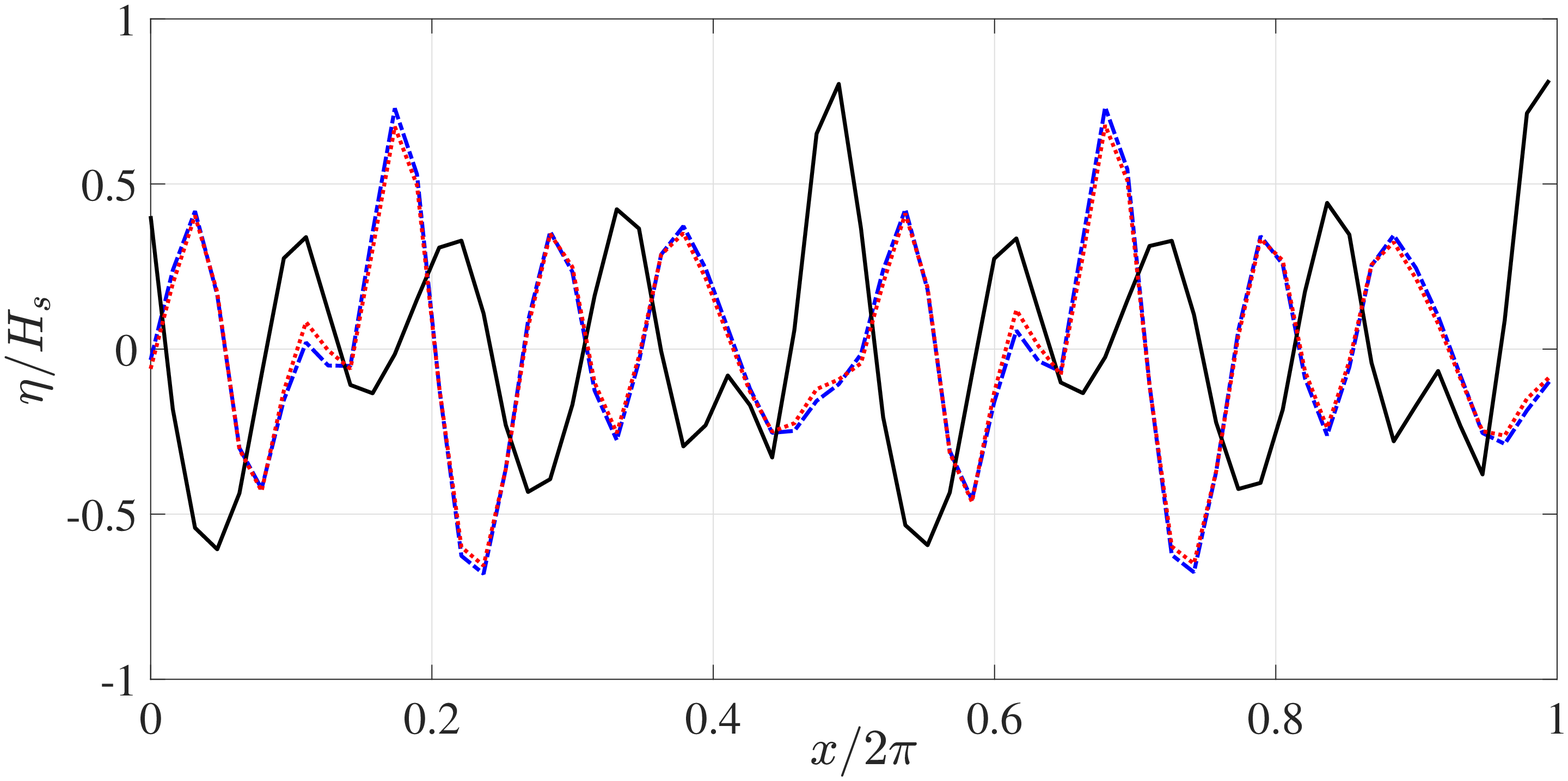}
         \caption{}
         \label{fig:wave2d_45}
     \end{subfigure}
     
        \caption{Surface elevations $\eta^{\text{true}}(x)$ (\hspace{-1.7mm}{\color{blue}\hdashrule[0.5ex]{0.7cm}{0.3mm}{1.3mm 0.5pt 0.5mm 0.5pt}}\hspace{-1.0mm}), $\eta^{\text{sim}}(x)$ with IEnKF-HOS-C ({\color{red}$\dotL$}) and HOS-C-only (\rule[0.5ex]{0.5cm}{1.0pt}) methods, at (a) $t/T_p=5$, (b) $t/T_p=25$, and (c) $t/T_p=45$, for the 2D wave field.}
        \label{fig:snapshots2d}
\end{figure}

\begin{figure}
     \centering
     \begin{subfigure}[b]{1\textwidth}
         \centering
         \includegraphics[trim=0cm 0cm 0cm 0cm, clip,scale=0.3]{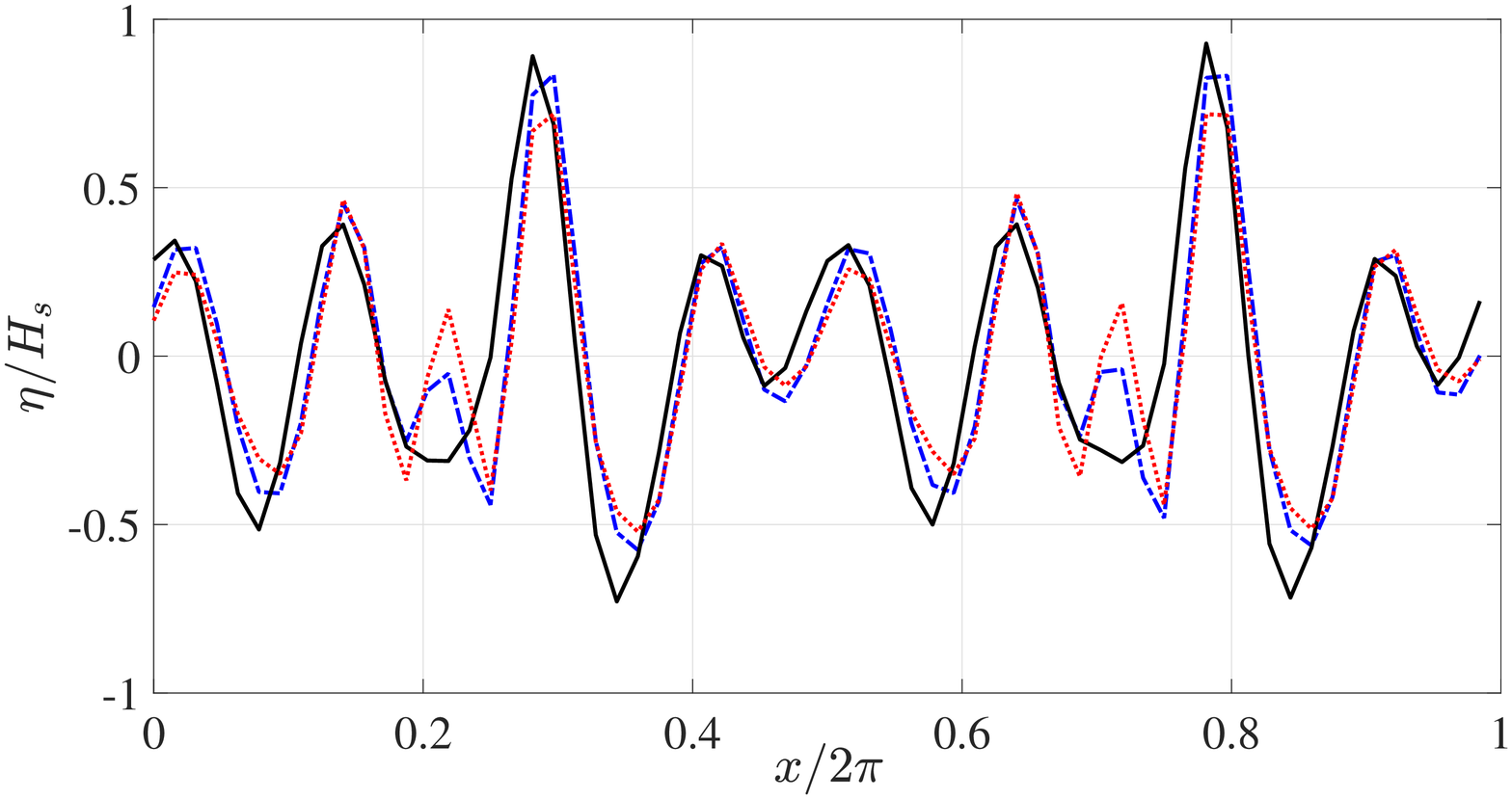}
         \caption{}
         \label{fig:wave3d_5}
     \end{subfigure}
     \begin{subfigure}[b]{1\textwidth}
         \centering
         \includegraphics[trim=0cm 0cm 0cm 0cm, clip,scale=0.3]{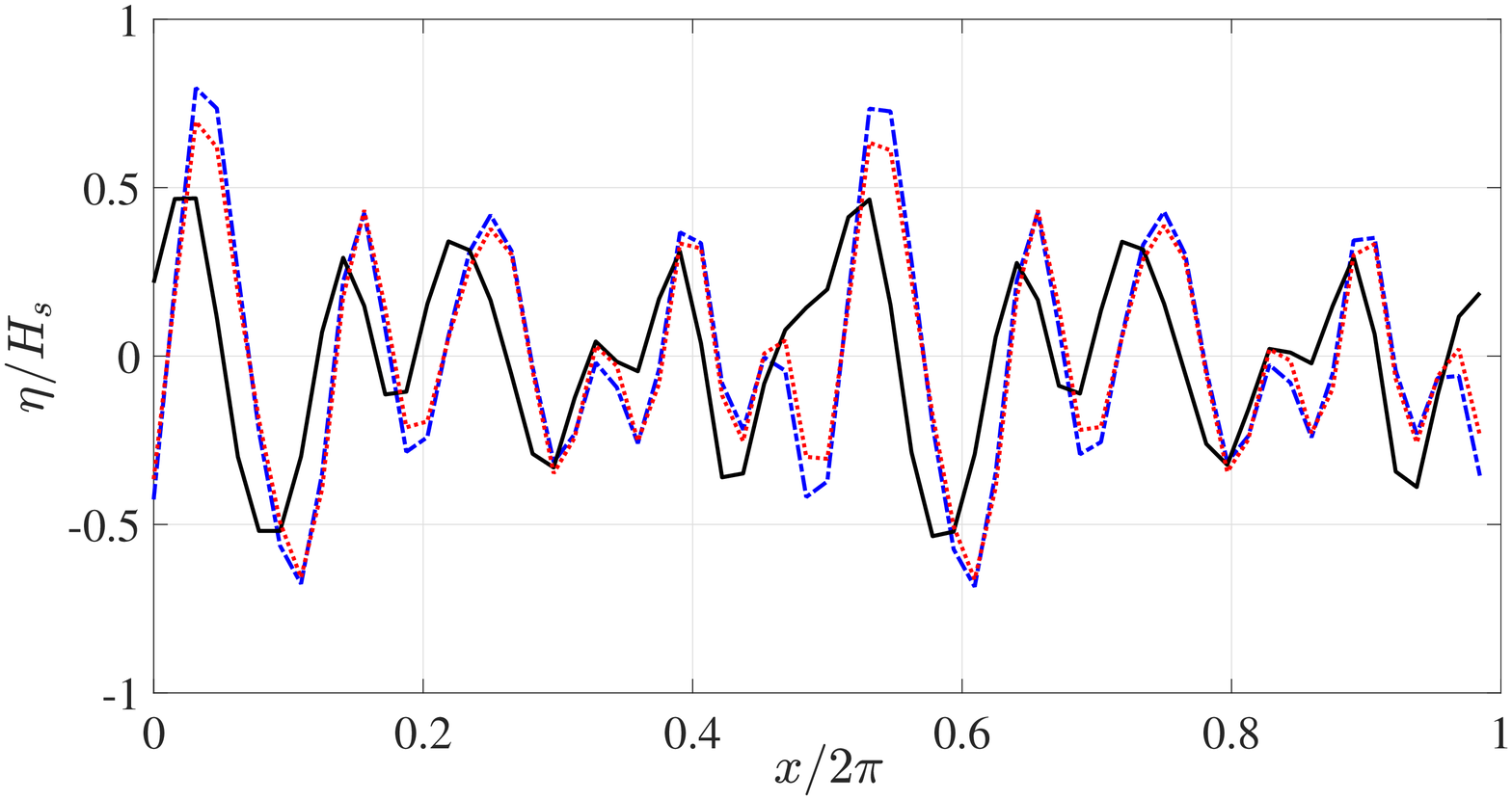}
         \caption{}
         \label{fig:wave3d_25}
     \end{subfigure}
     
     \begin{subfigure}[b]{1\textwidth}
         \centering
         \includegraphics[trim=0cm 0cm 0cm 0cm, clip,scale=0.3]{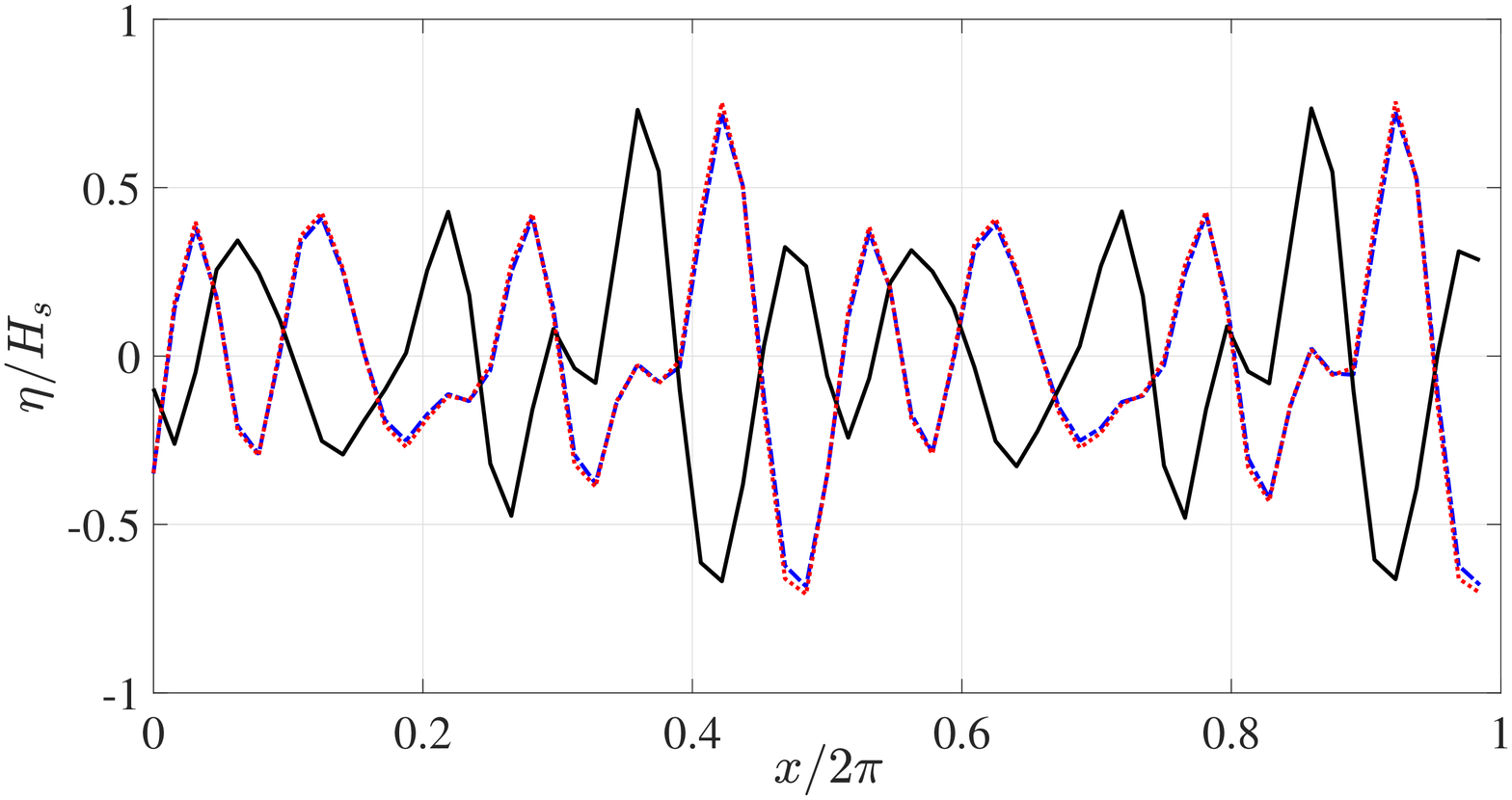}
         \caption{}
         \label{fig:wave3d_45}
     \end{subfigure}
     
        \caption{Surface elevations $\eta^{\text{true}}(x)$(\hspace{-1.5mm}{\color{blue}\hdashrule[0.5ex]{0.7cm}{0.3mm}{1.2mm 0.5pt 0.5mm 0.5pt}}\hspace{-1.0 mm}), $\eta^{\text{sim}}(x)$ with IEnKF-HOS-C ({\color{red}$\dotL$}) and HOS-C-only (\rule[0.5ex]{0.5cm}{1.0pt}) methods for the cross section $y/2\pi=0.3$ in the 3D wave field, at (a) $t/T_p=5$, (b) $t/T_p=25$, and (c) $t/T_p=45$.}
        \label{fig:snapshots3d}
\end{figure}
To further visualize the wave fields, figures~\ref{fig:snapshots2d}~and~\ref{fig:snapshots3d} show the snapshots of $\eta^{\text{true}}(x)$ and $\eta^{\text{sim}}(x)$ obtained from both
IEnKF-HOS-C and HOS-C-only methods at three time instants of $t/T_p=5,~25,~\text{and}~45$ for the 2D and 3D cases respectively. It can be found that, for both cases $\eta^{\text{sim}}(x)$ from the IEnKF-HOS-C method exhibits a much better agreement with $\eta^{\text{true}}(x)$ than that from the HOS-C-only method. At $t/T_p=45$, the IEnKF-HOS-C solution is already almost indistinguishable from $\eta^\text{true}(x)$.

\begin{figure}
     \centering
     \begin{subfigure}[b]{1\textwidth}
         \centering
         \includegraphics[trim=0cm 0cm 0cm 0cm, clip,scale=0.27]{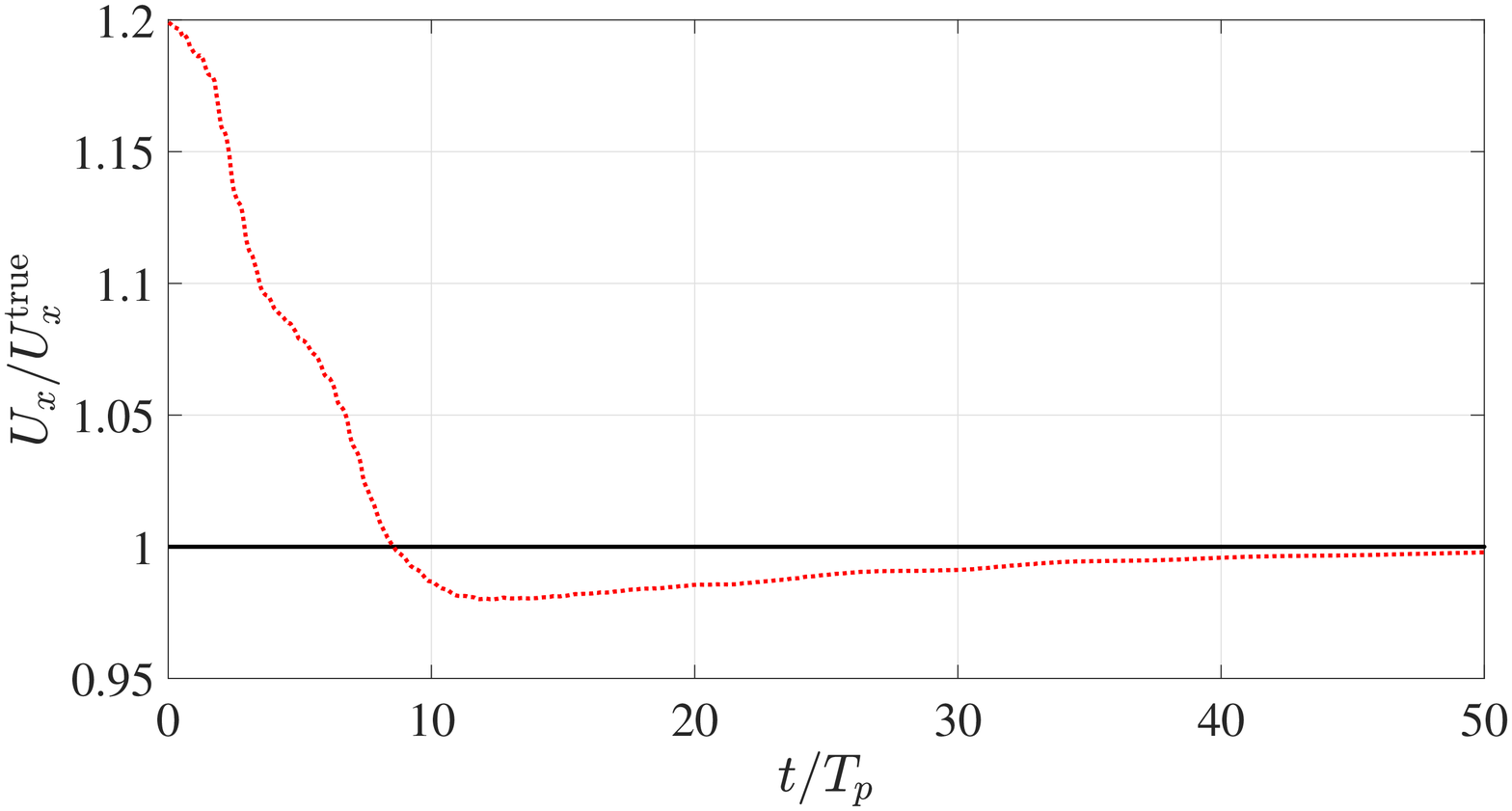}
         \caption{}
         \label{fig:u2d}
     \end{subfigure}
     \begin{subfigure}[b]{1\textwidth}
         \centering
         \includegraphics[trim=0cm 0cm 0cm 0cm, clip,scale=0.25]{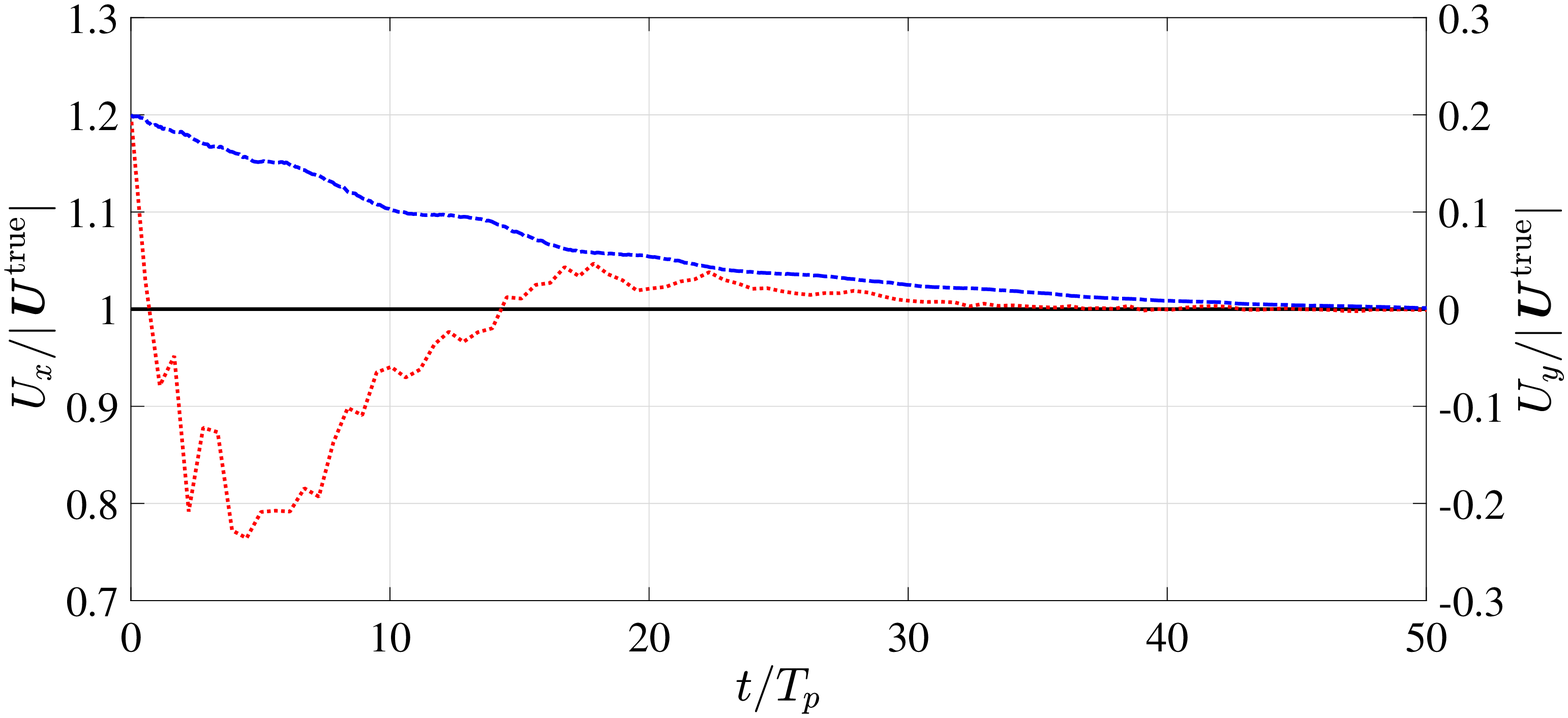}
         \caption{}
         \label{fig:u3d}
     \end{subfigure}
        \caption{$\boldsymbol{U}_{\text{a}}$ estimated by the IEnKF-HOS-C method (${U}_{x,\text{a}}$: {\color{red}\dotL}; ${U}_{y,\text{a}}$: \hspace{-1.3mm}{\color{blue}\hdashrule[0.5ex]{0.7cm}{0.3mm}{1.3mm 0.5pt 0.5mm 0.5pt}}\hspace{-1.2 mm}), in comparison with the true values ({\L}), for (a) 2D and (b) 3D cases with a steady and uniform current field. For (b) the $y$-component of the velocity is labeled on the right vertical axis.}
        \label{fig:u2d3d}
\end{figure}

Another important metric to evaluate the IEnKF-HOS-C performance is the estimated current velocity $\boldsymbol{U}_\text{a}$, which is shown in figure~\ref{fig:u2d3d} together with $\boldsymbol{U}^{\text{true}}$ for both 2D and 3D cases. It can be found that, as the measurements of the surface elevation are assimilated, $\boldsymbol{U}^\text{a}$ approaches $\boldsymbol{U}^\text{true}$ (although experiencing some fluctuations) with their difference practically negligible at $t/T_p=50$. This indicates that the IEnKF algorithm is successful in the estimation of current velocity in these cases. 

\begin{figure}
     \centering
     \begin{subfigure}[b]{1\textwidth}
         \centering
         \includegraphics[trim=0cm 0cm 0cm 0cm, clip,scale=0.26]{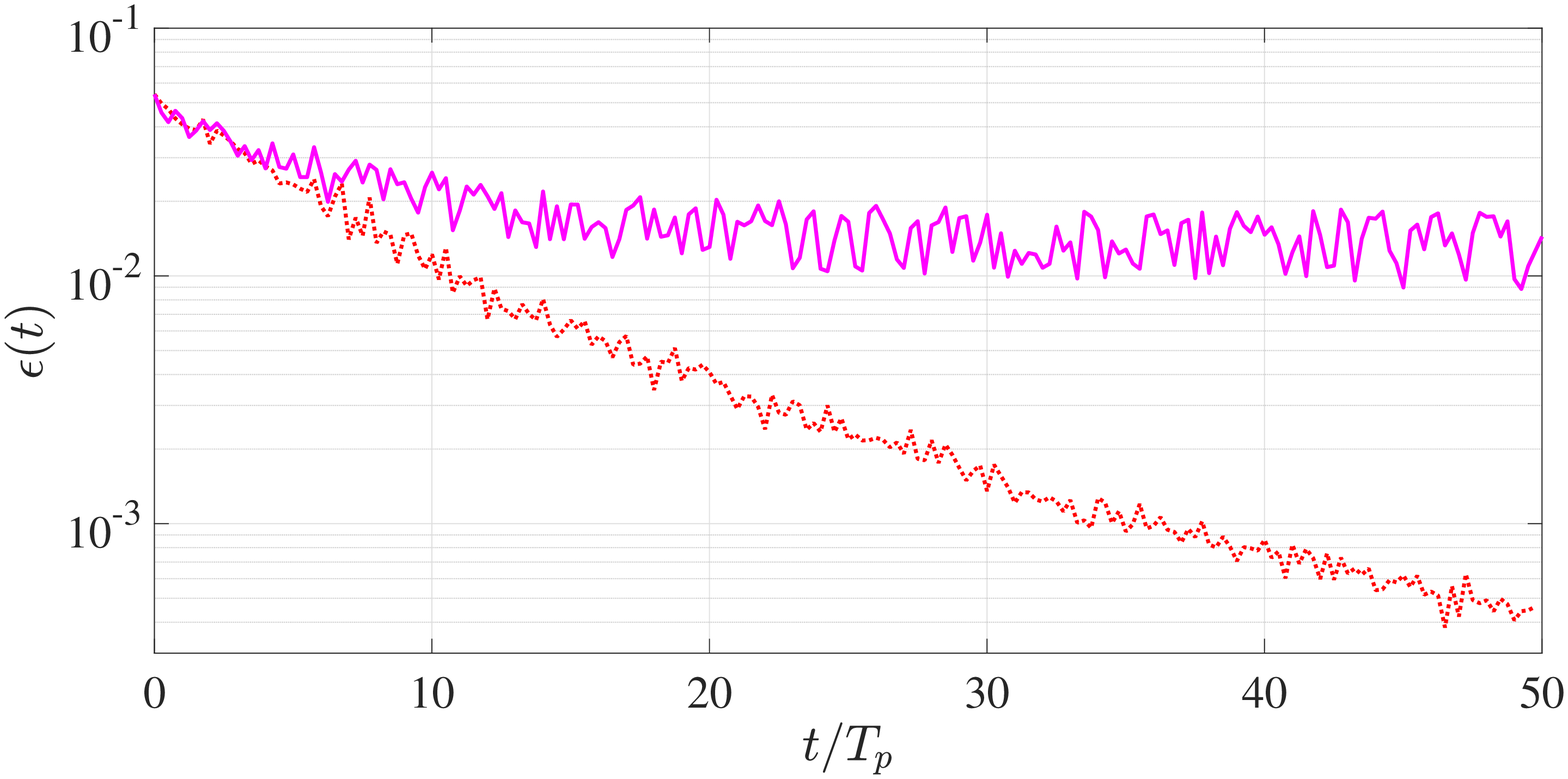}
         \caption{}
         \label{fig:e2dc}
     \end{subfigure}
     \begin{subfigure}[b]{1\textwidth}
         \centering
         \includegraphics[trim=0cm 0cm 0cm 0cm, clip,scale=0.26]{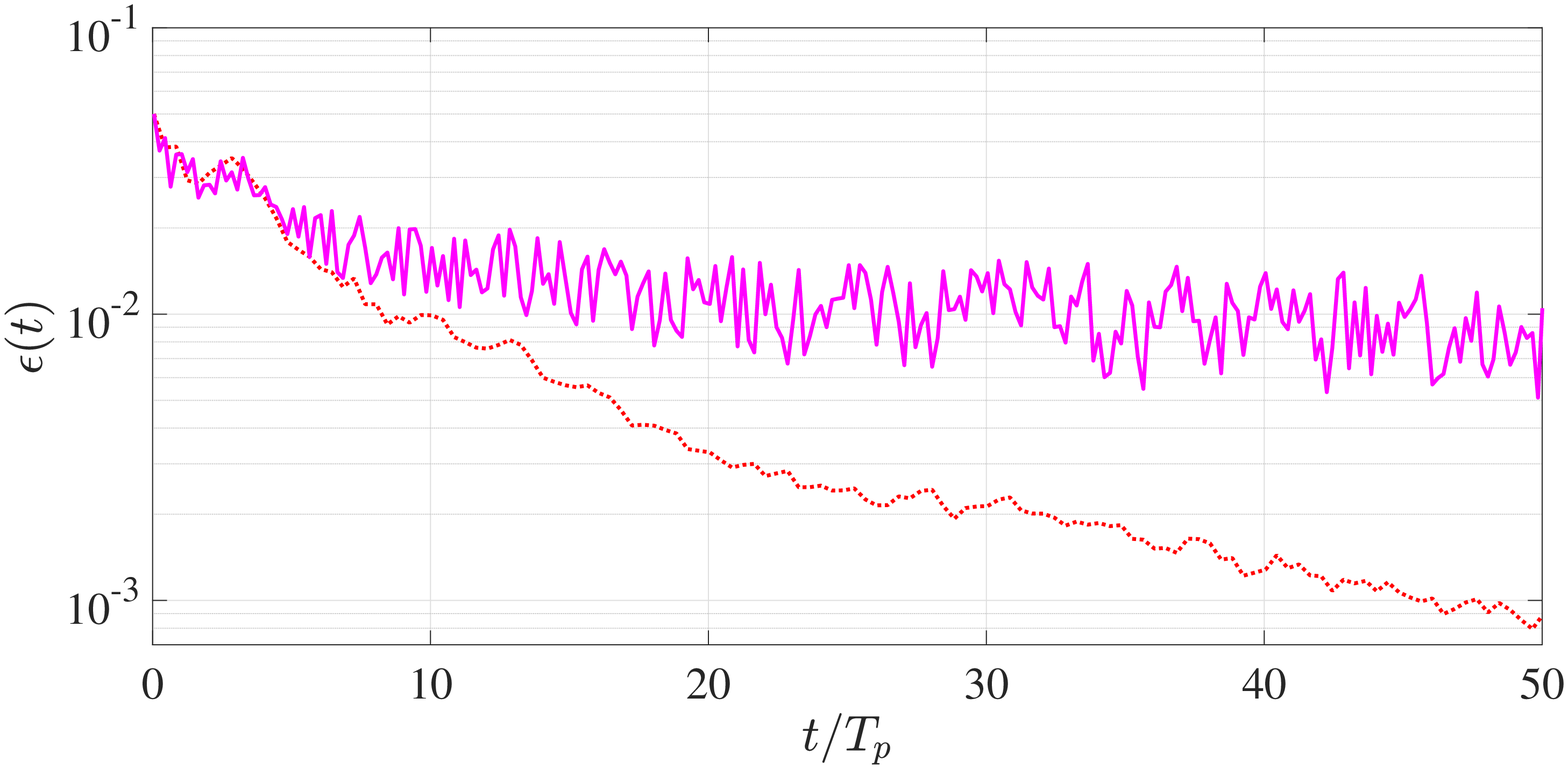}
         \caption{}
         \label{fig:e3dc}
     \end{subfigure}
        \caption{Errors $\epsilon(t)$ from IEnKF-HOS-C ({\color{red}\dotL}) and EnKF-HOS ({\color{magenta}\L}) methods, for the (a) 2D and (b) 3D cases with a steady and uniform current field.}
        \label{fig:e2d3dc}
\end{figure}
Finally, it is also of interest to compare the performance of IEnKF-HOS-C to the previous EnKF-HOS method developed by~\cite{wang2021phase}, where the latter characterizes the situation of a biased physical model coupled with DA (that can partially correct the solution).  Figure~\ref{fig:e2d3dc} plots the errors $\epsilon(t)$ obtained from the IEnKF-HOS-C and EnKF-HOS methods for both 2D and 3D cases. We observe that $\epsilon(t)$ from EnKF-HOS decreases in time with a much slower rate than that from IEnKF-HOS-C, and reaches a constant level after $t\approx 15T_p$ (representing a balance between prediction model error growth and DA correction). It is clear that the IEnKF-HOS-C performs much better, with its error one order of magnitude smaller than that from EnKF-HOS at $t=50T_p$.

\vspace{0.5cm}
\begin{description}
   \item[Unsteady cases]
\end{description}

\begin{figure}
     \centering
     \begin{subfigure}[b]{1\textwidth}
         \centering
         \includegraphics[trim=0cm 0cm 0cm 0cm, clip,scale=0.25]{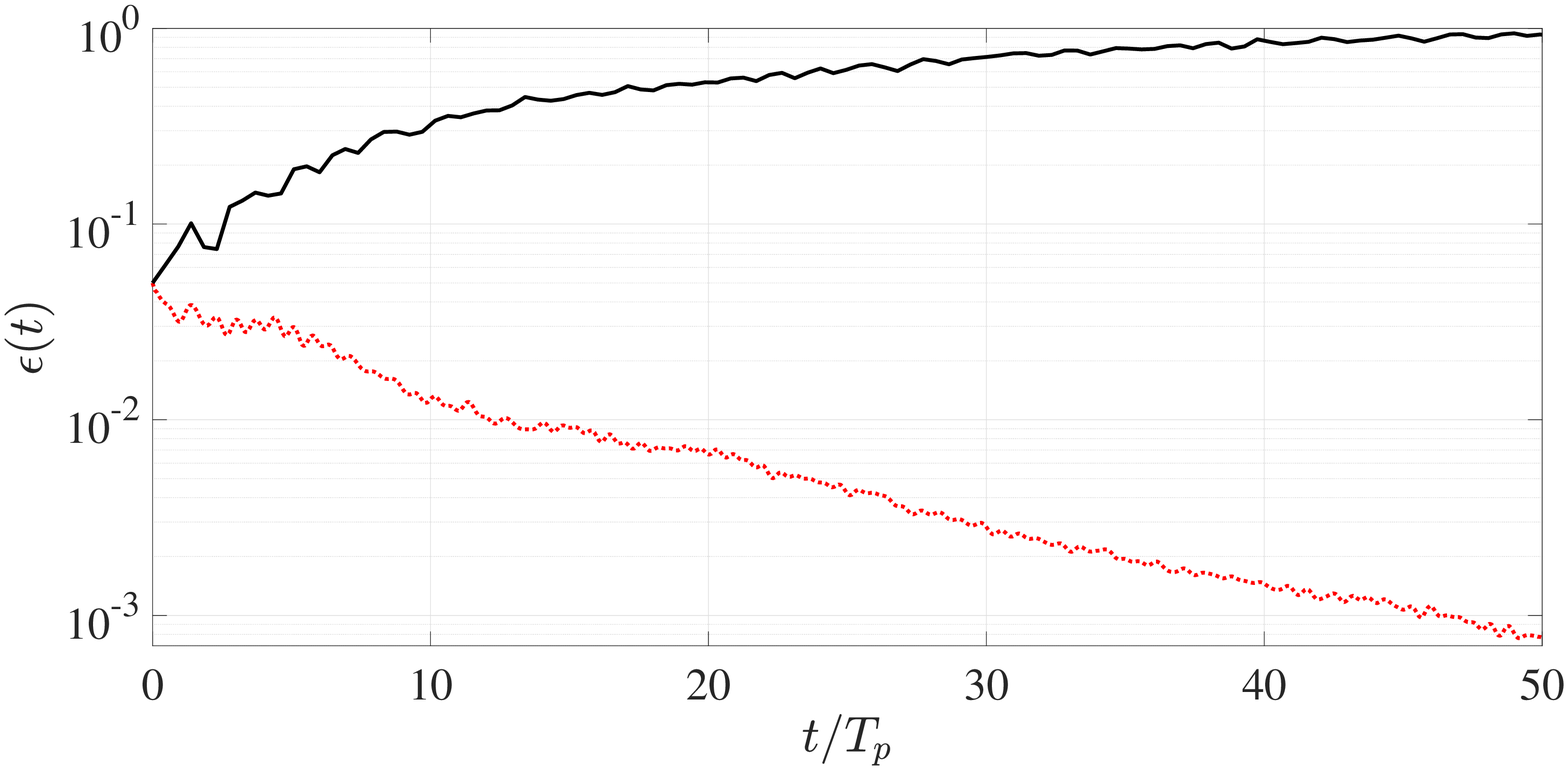}
         \caption{}
         \label{fig:elinear}
     \end{subfigure}
     \begin{subfigure}[b]{1\textwidth}
         \centering
         \includegraphics[trim=0cm 0cm 0cm 0cm, clip,scale=0.25]{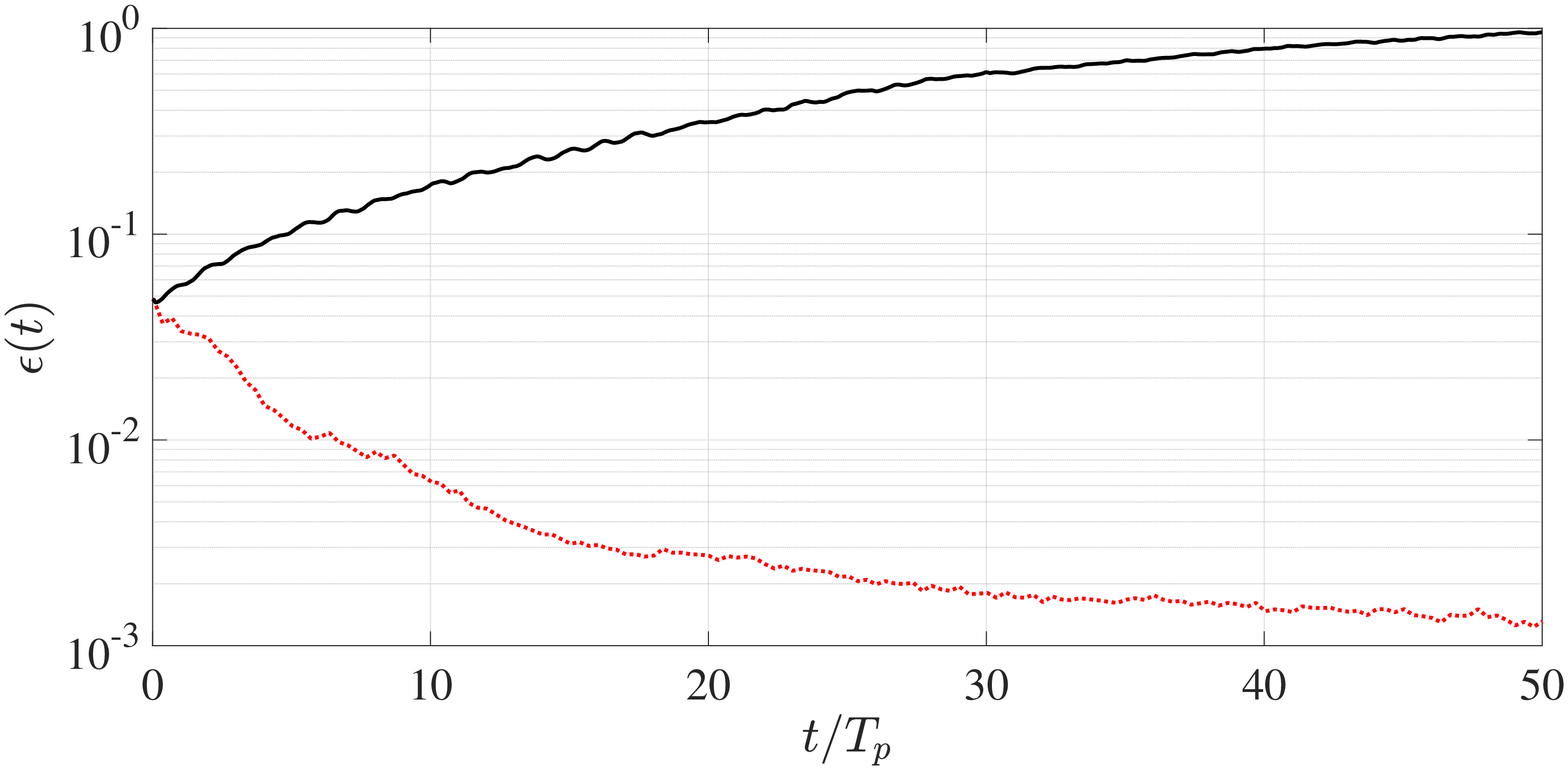}
         \caption{}
         \label{fig:ecos}
     \end{subfigure}
        \caption{Errors $\epsilon(t)$ from IEnKF-HOS-C ({\color{red}\dotL}) and HOS-C-only (\rule[0.5ex]{0.5cm}{1.0pt}) methods for the cases with unsteady and uniform current fields: (a)~\eqref{eq:linear} and (b)~\eqref{eq:cos}.}
        \label{fig:elc}
\end{figure}

\begin{figure}
     \centering
     \begin{subfigure}[b]{1\textwidth}
         \centering
         \includegraphics[trim=0cm 0cm 0cm 0cm, clip,scale=0.25]{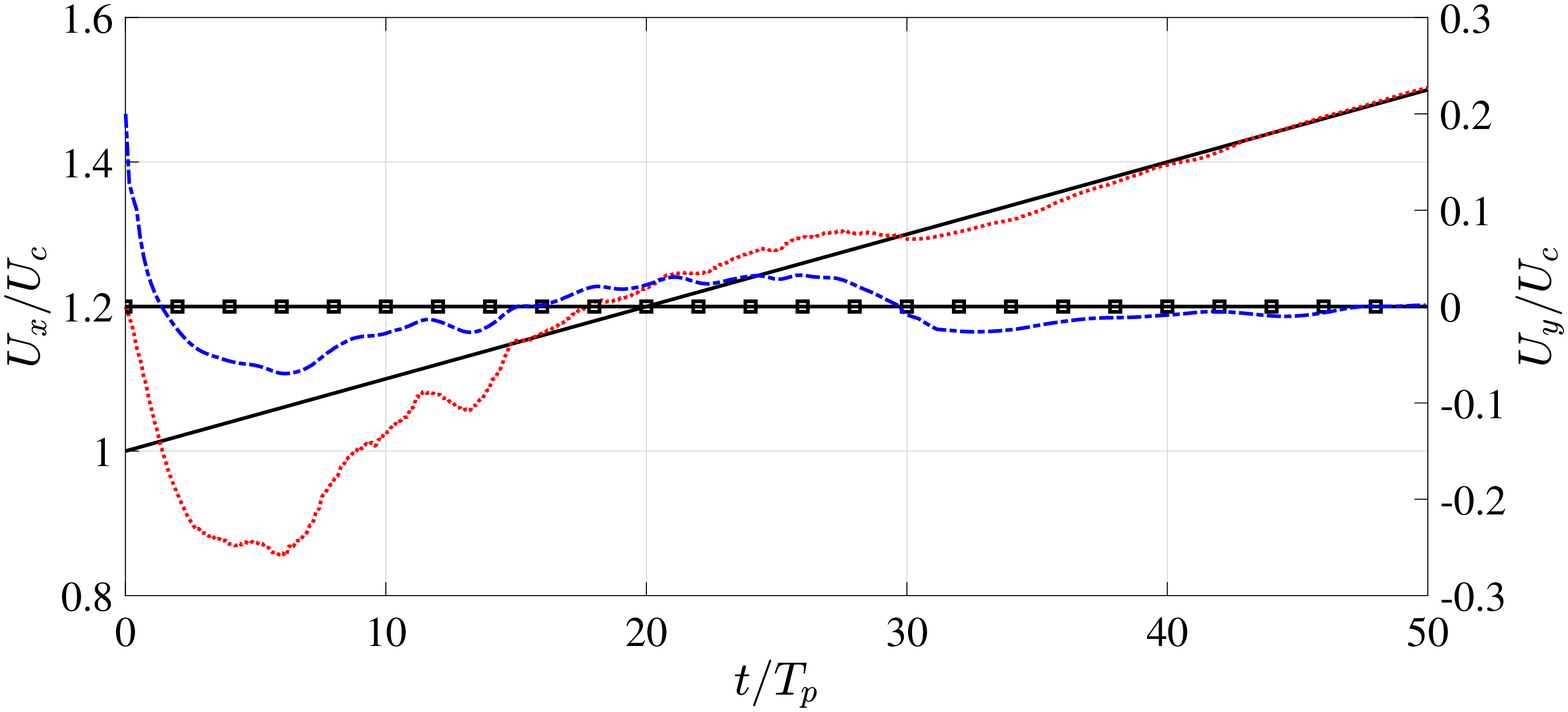}
         \caption{}
         \label{fig:ulinear}
     \end{subfigure}
     \begin{subfigure}[b]{1\textwidth}
         \centering
         \includegraphics[trim=0cm 0cm 0cm 0cm, clip,scale=0.25]{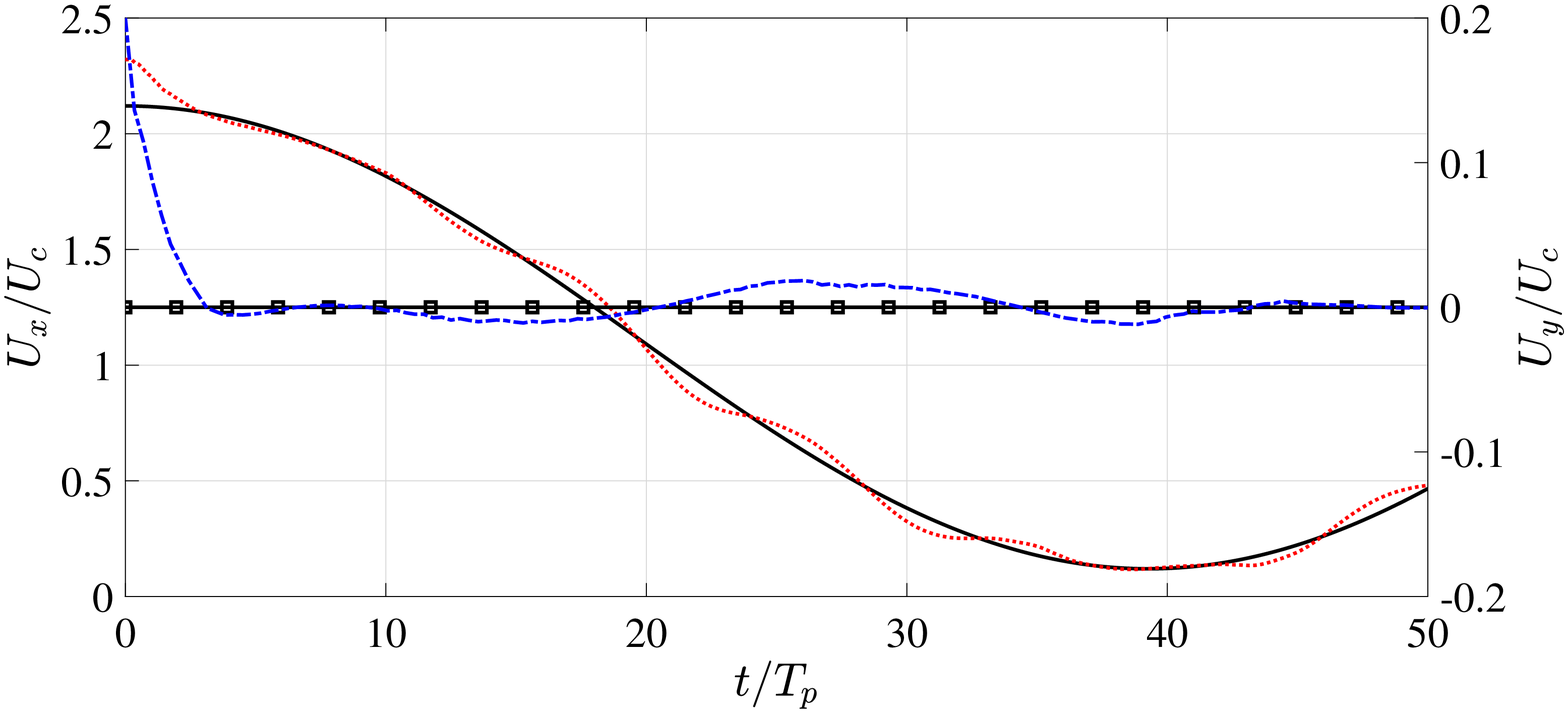}
         \caption{}
         \label{fig:ucos}
     \end{subfigure}
        \caption{Estimations $\boldsymbol{U}_{\text{a},j}$ from the IEnKF-HOS-C method ($U_{x,\text{a}}$: {\color{red}\dotL};~$U_{y,\text{a}}: ${\color{blue}{\hdashrule[0.5ex]{0.7cm}{0.3mm}{1.3mm 0.5pt 0.4mm 0.5pt}}\hspace{-1.0mm}}), in comparison with the true values ($U_{x}^{\text{true}}(t)$: {\L};~$U_{y}^{\text{true}}(t)=0$:\Lbox), for the cases with unsteady and uniform current fields: (a)~\eqref{eq:linear} and (b)~\eqref{eq:cos}.}
        \label{fig:uunsteady}
\end{figure}
In this section, we test the performance of the IEnKF-HOS-C method for unsteady and uniform current fields. We focus on 3D wave fields hereafter and for this section we consider both linear and sinusoidal time variations of the current fields, described by
\begin{equation}
    U_x^{\text{true}}(t)=(1+\alpha_1 t/T_p)U_c,
    \label{eq:linear}
\end{equation}
and
\begin{equation}
    U_x^{\text{true}}(t)=\left(\cos(\alpha_2 t/T_p)+1.12\right)U_c,
    \label{eq:cos}
\end{equation}
with $\alpha_1=0.01$, $\alpha_2=0.08$, and $U_c=0.25v_p(k_p)$. The initial guess of current velocity is also prescribed by~\eqref{eq:uy}, based on which the ensemble is then produced with $c_u=0.04(U_c)^2$ and $a_u \rightarrow \infty$. Noisy measurements of the wave field at $24$ randomly sampled locations are assimilated with an interval $\tau=T_p/32$. 

Figure~\ref{fig:elc} shows the errors $\epsilon(t)$ from the IEnKF-HOS-C and HOS-C-only methods, for the two current fields \eqref{eq:linear} and \eqref{eq:cos}. For the HOS-C-only simulations, $\epsilon(t)$ grows quickly in time (somewhat faster than the steady current cases at the early stage of simulations) and reaches $\mathcal{O}(1)$ at $t/T_p \approx 40 $. The IEnKF-HOS-C is again successful in accurately estimating the wave fields with the error $\epsilon(t)$ reducing to $\mathcal{O}(10^{-3})$ at  $t/T_p \approx 50 $.

Figure~\ref{fig:uunsteady} presents the estimated velocity by the IEnKF-HOS-C method, in terms of its evolution in time and comparison to the true current velocity $\boldsymbol{U}^{\text{true}}(t)$. For both types of current fields \eqref{eq:linear} and \eqref{eq:cos}, the IEnKF-HOS-C method is able to track the variation of the current field, with $\boldsymbol{U}_{\text{a},j}$ converging to the vicinity of the true time series after $5\sim 15 T_p$. We remark that the capability to capture the unsteadiness of the current is achieved through the IEnKF procedure, in spite of the persistence model \eqref{eq:persis} in the forecast step.

\subsubsection{Results for non-uniform current fields}

We further demonstrate the effectiveness of the IEnKF-HOS-C method for the 3D wave field evolution under the effect of non-uniform current fields. The results below are presented for steady and unsteady non-uniform current fields respectively. 

\vspace{0.5cm}
\begin{description}
   \item[Steady case]
\end{description}

\begin{figure}
  \centerline{\includegraphics[scale =0.3]{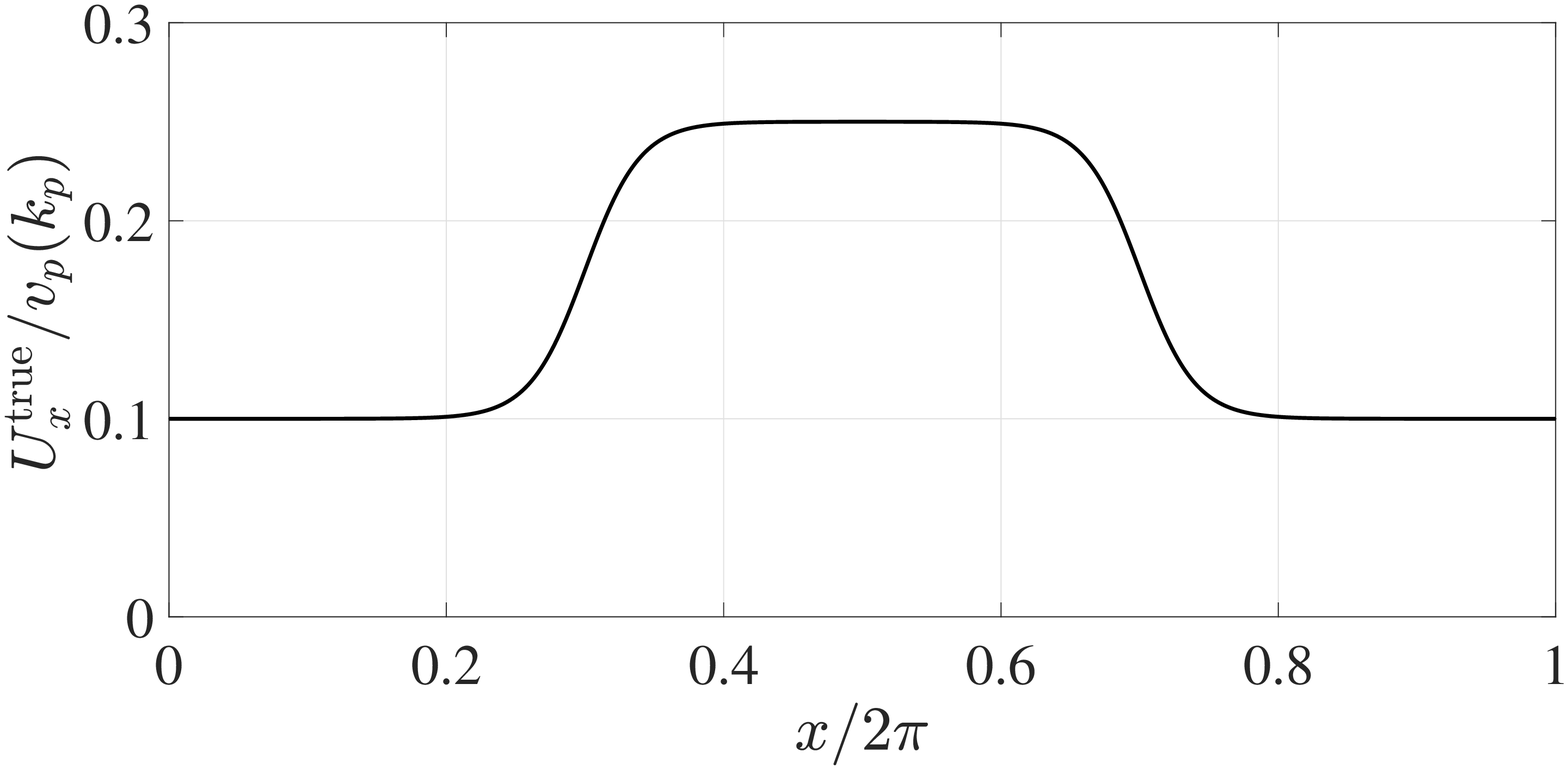}}
  \caption{$U_x^{\text{true}}(x)$ as the true steady and non-uniform current field described by \eqref{eq:nucur}.}
\label{fig:ux2}
\end{figure}

We consider the wave field evolution under the effect of a steady and non-uniform (varying in $x$ direction) current field, which is described by
\begin{equation}
    U^{\text{true}}_x(x)=
    \begin{cases}
    \displaystyle\frac{U_1+e^{\gamma x-q}U_2}{e^{\gamma x-q}+1},~&\text{for}~\displaystyle0\leq x\leq \pi \\
    U_x^{\text{true}}(2\pi-x),~&\text{for}~\displaystyle \pi < x\leq 2\pi
    \end{cases}
    \label{eq:nucur}
\end{equation}
and plotted in figure~\ref{fig:ux2}. The parameter values are chosen as $U_1=0.1v_p(k_p)$, $U_2=0.25v_p(k_p)$, $q=15$, and $\gamma=\displaystyle\frac{25}{\pi}$ such that the transitions between the locally uniform regions are smooth (i.e., slow variation of the current field compared to the wave oscillation, and thus compatible with \eqref{eq:bc1} and \eqref{eq:bc2}).

In this case, we set the initial guess of current velocity to be uniform as before (assuming no spatial distribution information is accessible \emph{a priori}):
\begin{equation}
    U_{x,0}(\boldsymbol{x})=0.15v_p(k_p),
\label{eq:uniformigx}
\end{equation}
\begin{equation}
 U_{y,0}(\boldsymbol{x})=0.02v_p({k}_p),
\label{eq:uniformigy}
\end{equation}
with the initial ensemble for both velocity components generated by \eqref{eq:ustar} with $c_u=0.04(U_1)^2$ and $a_u=\pi/2$.

\begin{figure}
  \centerline{\includegraphics[scale =0.3]{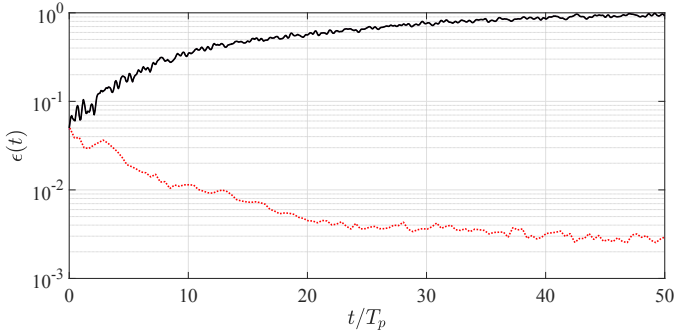}}
  \caption{Errors $\epsilon(t)$ from the IEnKF-HOS-C ({\color{red}\dotL}) and HOS-C-only (\rule[0.5ex]{0.5cm}{1.0pt}) methods for the current field described by \eqref{eq:nucur}.}
\label{fig:e_non_uniform_steady}
\end{figure}

With the IEnKF-HOS-C method, data of surface elevation at $d=64$ randomly selected locations are assimilated into the numerical model with an interval $\tau=T_p/64$. Figure~\ref{fig:e_non_uniform_steady} plots the errors $\epsilon(t)$ obtained from the IEnKF-HOS-C and HOS-C-only methods for this case, showing again that IEnKF-HOS-C is successful in estimating the wave states, in contrast to the HOS-C-only simulation.

\begin{figure}
  \centerline{\includegraphics[scale =0.6]{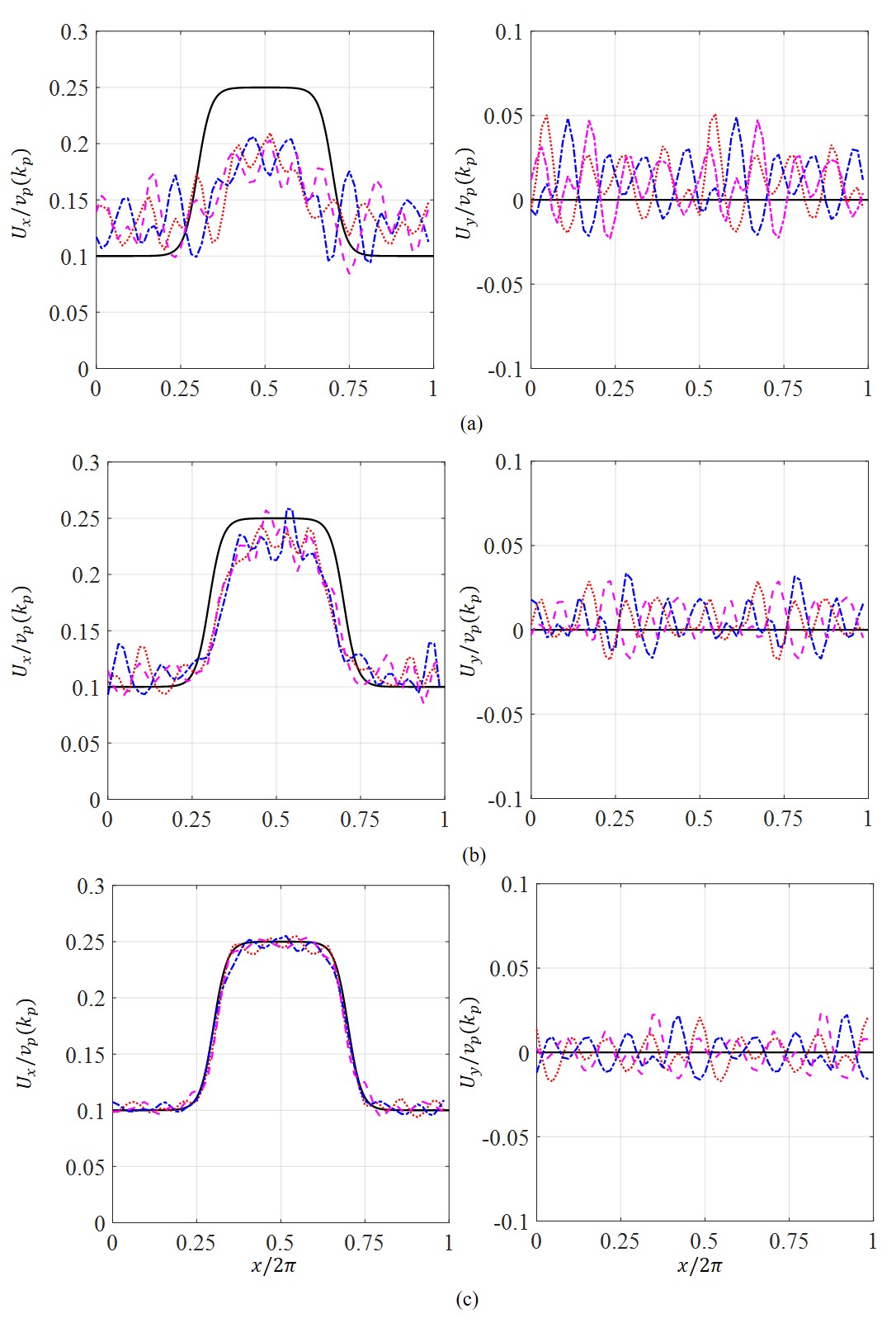}}
  \caption{Estimations $U_{x,\text{a}}$ (left) and $U_{y,\text{a}}$ (right) by IEnKF-HOS-C at (a) $t/T_p=5$, (b) $t/T_p=25$, and (c) $t/T_p=45$ in comparison with the true current field \eqref{eq:nucur} (\L), for three cross sections: $y/2\pi=0.25$({\color{red}{\dotL}}), $y/2\pi=0.50$(\hspace{-1.0mm}{\color{blue}\hdashrule[0.5ex]{0.7cm}{0.3mm}{1.2mm 0.5pt 0.5mm 0.5pt}}\hspace{-2.0mm}), and $y/2\pi=0.75$({\color{magenta}\dashL}).}
\label{fig:non_uniform_velocity_v2}
\end{figure}

The current fields captured by IEnKF-HOS-C, in terms of the snapshots of $U_{x,\text{a}}(\boldsymbol{x})$ and $U_{y,\text{a}}(\boldsymbol{x})$ at three cross sections of constant $y$ for $t/T_p=5,~25,~\text{and}~45$, are plotted in figures~\ref{fig:non_uniform_velocity_v2}. We see that the estimated velocity starts from a constant value and converges to the true field (with its variation in $x$ captured at all cross sections) as the time increases. This successful estimation of the current field is the basis for the accurate prediction of the wave states seen in figure \ref{fig:e_non_uniform_steady}.

\vspace{0.5cm}
\begin{description}
   \item[Unsteady cases]
\end{description}
The climax of the synthetic cases is the application of the IEnKF-HOS-C method to the wave field evolution impacted by a current field featuring both spatial and temporal variations. In particular, we assign a sinusoidal variation to the current field \eqref{eq:nucur} to produce the true current velocity field:
\begin{equation}
    U^{\text{true}}_x(\boldsymbol{x},t)=
    \begin{cases}
    \displaystyle\frac{U_1+e^{\gamma x-q}U_2}{e^{\gamma x-q}+1}\cos(\alpha_3 t/T_p),~&\text{for}~\displaystyle0\leq x\leq \pi \\
    U_x^{\text{true}}(2\pi-x),~&\text{for}~\displaystyle \pi < x\leq 2\pi
    \end{cases}.
    \label{eq:nuuncur}
\end{equation}
where $\alpha_3=0.02$. For this case we use the same initial guess of current velocity as in the steady case, i.e.~\eqref{eq:uniformigx} and~\eqref{eq:uniformigy}, as well as the same corresponding initial ensembles generated by \eqref{eq:ustar} with $c_u=0.04(U_1)^2$ and $a_u=\pi/2$.

\begin{figure}
  \centerline{\includegraphics[scale =0.3]{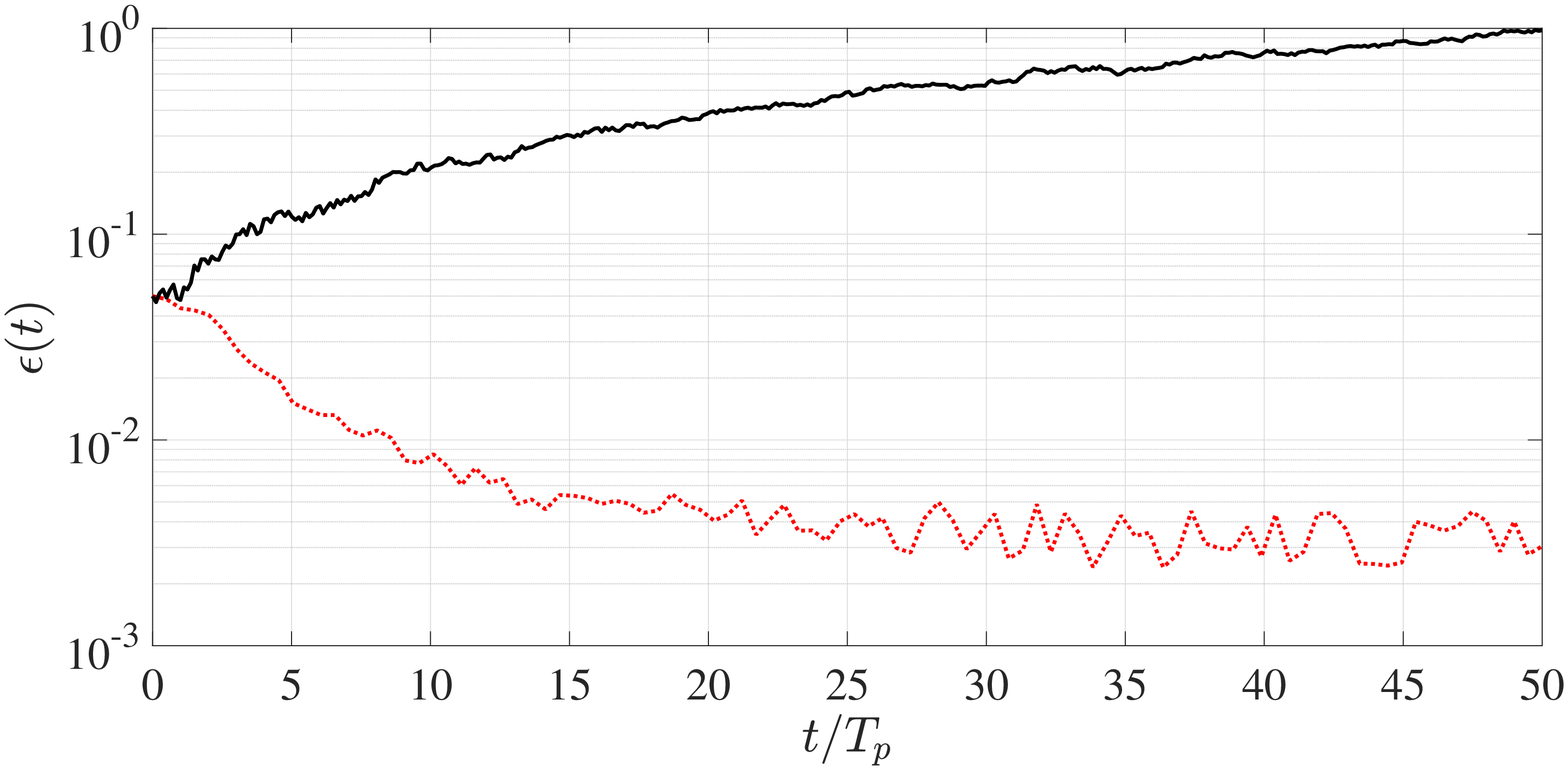}}
  \caption{Error $\epsilon(t)$ from the IEnKF-HOS-C ({\color{red}\dotL}) and HOS-C-only (\rule[0.5ex]{0.5cm}{1.0pt}) methods for the current field described by \eqref{eq:nuuncur}.}
\label{fig:e_non_uniform_unsteady}
\end{figure}

\begin{figure}
  \centerline{\includegraphics[scale =0.6]{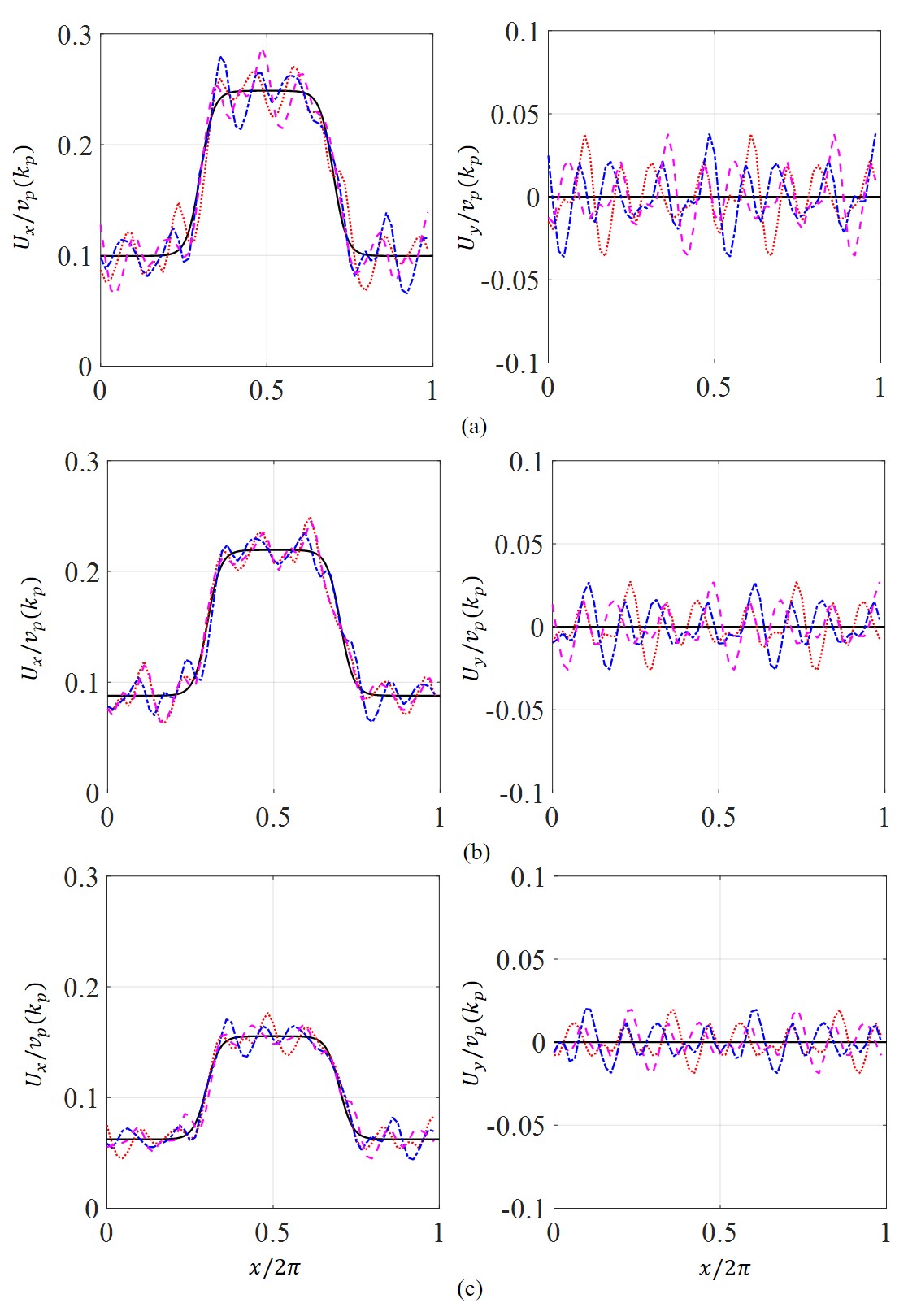}}
  \caption{Estimations $U_{x,\text{a}}$ (left) and $U_{y,\text{a}}$ (right) by IEnKF-HOS-C at (a) $t/T_p=5$, (b) $t/T_p=25$, and (c) $t/T_p=45$ in comparison with the true field \eqref{eq:nuuncur} (\L), for three cross sections: $y/2\pi=0.25$({\color{red}{\dotL}}), $y/2\pi=0.50$(\hspace{-1.2mm}{\color{blue}\hdashrule[0.5ex]{0.7cm}{0.3mm}{1.3mm 0.5pt 0.5mm 0.5pt}}\hspace{-0.8mm}), and $y/2\pi=0.75$({\color{magenta}\dashL}).}
\label{fig:non_uniform_unsteady_velocity}
\end{figure}

The time series of $\epsilon(t)$ from the IEnKF-HOS-C and HOS-C-only methods are shown in figure \ref{fig:e_non_uniform_unsteady}, which demonstrates, similar to all above cases, the effectiveness of IEnKF-HOS-C in reproducing the wave states. The estimated current field by IEnKF-HOS-C is further plotted in figure~\ref{fig:non_uniform_unsteady_velocity}, in terms of the snapshots of $U_{x,\text{a}}$ and $U_{y,\text{a}}$ at the three cross sections of constant $y$  for three time instants $t/T_p=5,~25,~\text{and}~45$. Starting from an initial guess of a constant field, $U_{x,\text{a}}$ captures both the spatial and temporal variation of $U^{\text{true}}_x(\boldsymbol{x},t)$ and $U_{y,\text{a}}$ approaches zero uniformly in time with sequential data assimilated to the algorithm.

\subsection{The case with real wave data}

\begin{figure}
  \centerline{\includegraphics[trim=0cm 0cm 0cm 0cm, clip,scale=0.6]{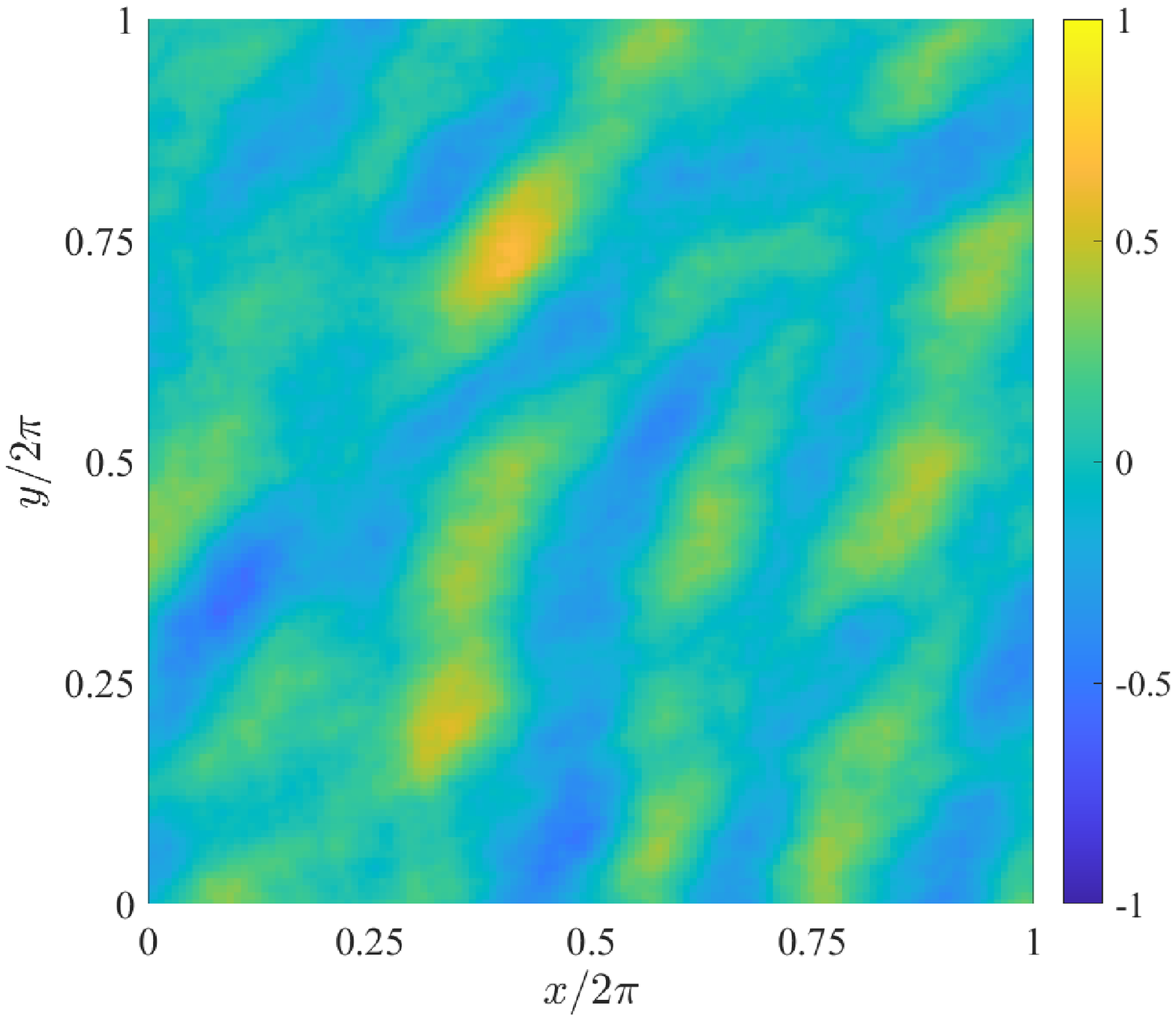}}
  \caption{Initial surface elevation $\eta_{m,0}(\boldsymbol{x})/H_s$ (with $H_s=1.50\text{m}$) measured by radar at $t=t_0$, i.e., $23:18:32$ UTC on 09/17/2013.}
\label{fig:ic_eta_real}
\end{figure}

In what follows, we test the IEnKF-HOS-C method using real measurements of the ocean wave field presented in ~\cite{lyzenga2015real}. The measurements are obtained from an onboard $25\text{kW}$ X-band ($9.4~\text{GHz}$) Doppler coherent marine radar off the coast of southern California. A patch of the radar-scanned area, which is fixed in the local radar coordinate system and covers a 480m$\times$480m region, is selected as the domain of interest. The numerical simulation starts from $23:18:32$ UTC on 09/17/2013 and lasts for $40T_p$ with $T_p=11.28\text{s}$. The initial condition is taken from the measurements in the computational domain (figure \ref{fig:ic_eta_real}) featuring a global wave steepness $k_pH_s/2=0.02$. We use $64\times64$ grid points in the simulation, which is consistent to the resolution in the radar data set. The DA interval is set to be the same as the data acquisition interval of the radar, which fluctuates around $T_p/4=2.82\text{s}$.

The reference velocity of the current field in this case is taken from the track of an \emph{in situ} floating buoy, with its location known at the beginning and end of the simulation time interval. While it is preferable to obtain the spatial and temporal variations of the current field, the available information only allows us to compute the mean velocity of the floating buoy, as $\boldsymbol{U}_b=(0.2958,-0.3944)~\text{m/s}$. On the other hand, due to the relatively small size of the simulation domain and short time interval, it can be justified to consider a uniform and steady current field with velocity specified by $\boldsymbol{U}_b$.

To generate ensemble of measurements of the wave field, we use~\eqref{eq:etaini} and \eqref{eq:lpsi} with $c_w=0.0025\sigma_{\eta}^2$ and $a_w=120\text{m}$ defined in~\eqref{eq:noise1}. Four different initial guesses of the current velocity are considered, including (a) $\boldsymbol{U}_0=(0.1958,-0.4944)$; (b) $\boldsymbol{U}_0=(0.3958, -0.2944)$; (c) $\boldsymbol{U}_0=(0.3958, -0.4944)$; and (d) $\boldsymbol{U}_0=(0.1958, -0.2944)$, all representing a shift of $0.1$m/s in different directions of each component of $\boldsymbol{U}_b$. The ensemble of the initial current velocity is generated by \eqref{eq:ustar} with $c_u=0.02\text{m}^2/\text{s}^2$ and $a_u=120\text{m}$.

An issue that needs to be considered for such a realistic case with uncertain boundary conditions is the problem of predictable zone and its potential mismatch with the measurement region. This issue has been discussed in detail in \cite{wang2021phase}, which involves the application of a modified analysis equation in EnKF (IEnKF in this case). For conciseness, we do not present the details again in this paper but refer the interested readers to our previous paper \cite{wang2021phase}. These previously developed techniques are also applied here, with the only additional complexity that the  estimated current velocity $\boldsymbol{U}_{\text{a},j}$ now needs to be added to the wave group velocity (accounting for the Doppler shift) to determine the boundaries of the predictable zones.  

\begin{figure}
     \centering
     \begin{subfigure}[b]{1\textwidth}
         \centering
         \includegraphics[trim=0cm 0cm 0cm 0cm, clip,scale=0.29]{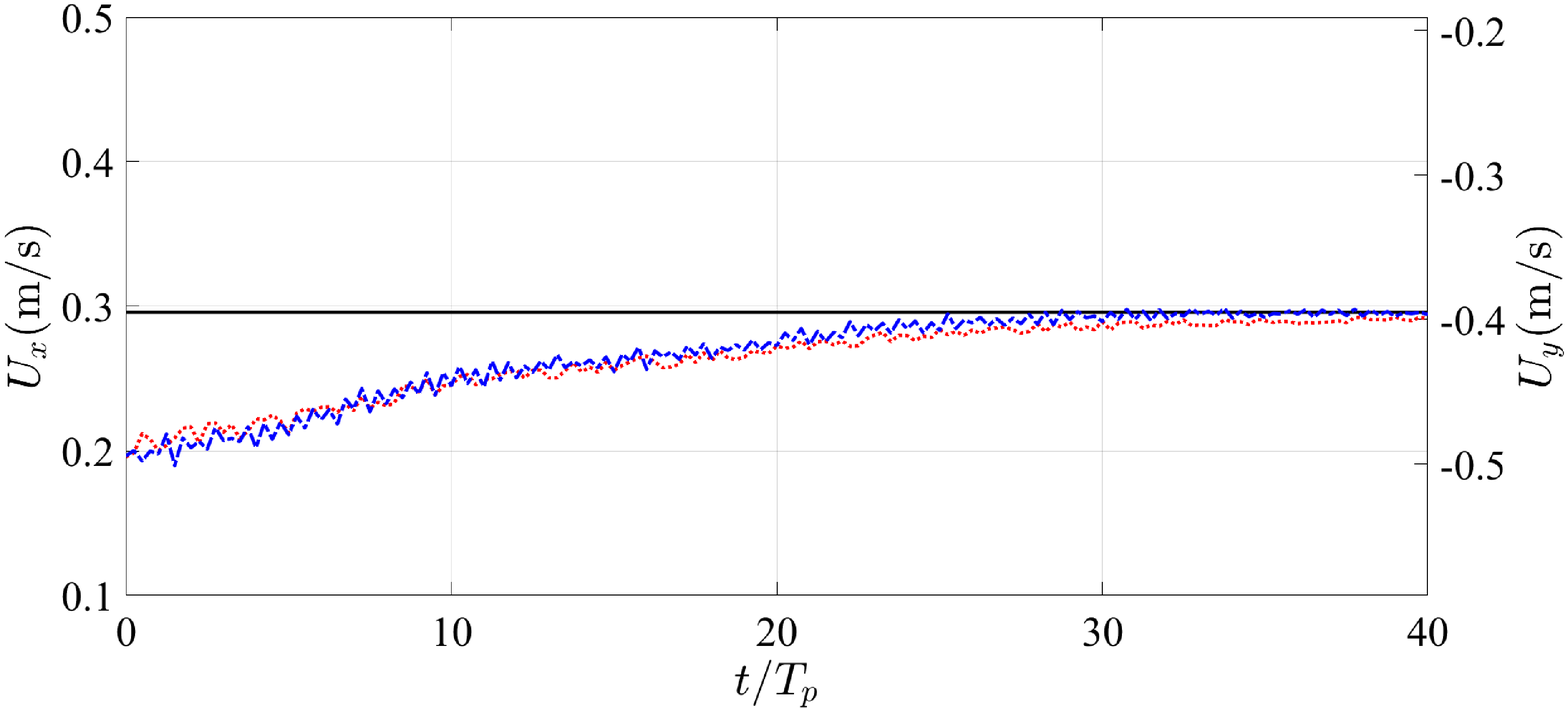}
         \caption{}
         \label{fig:ureal1}
     \end{subfigure}
     \begin{subfigure}[b]{1\textwidth}
         \centering
         \includegraphics[trim=0cm 0cm 0cm 0cm, clip,scale=0.29]{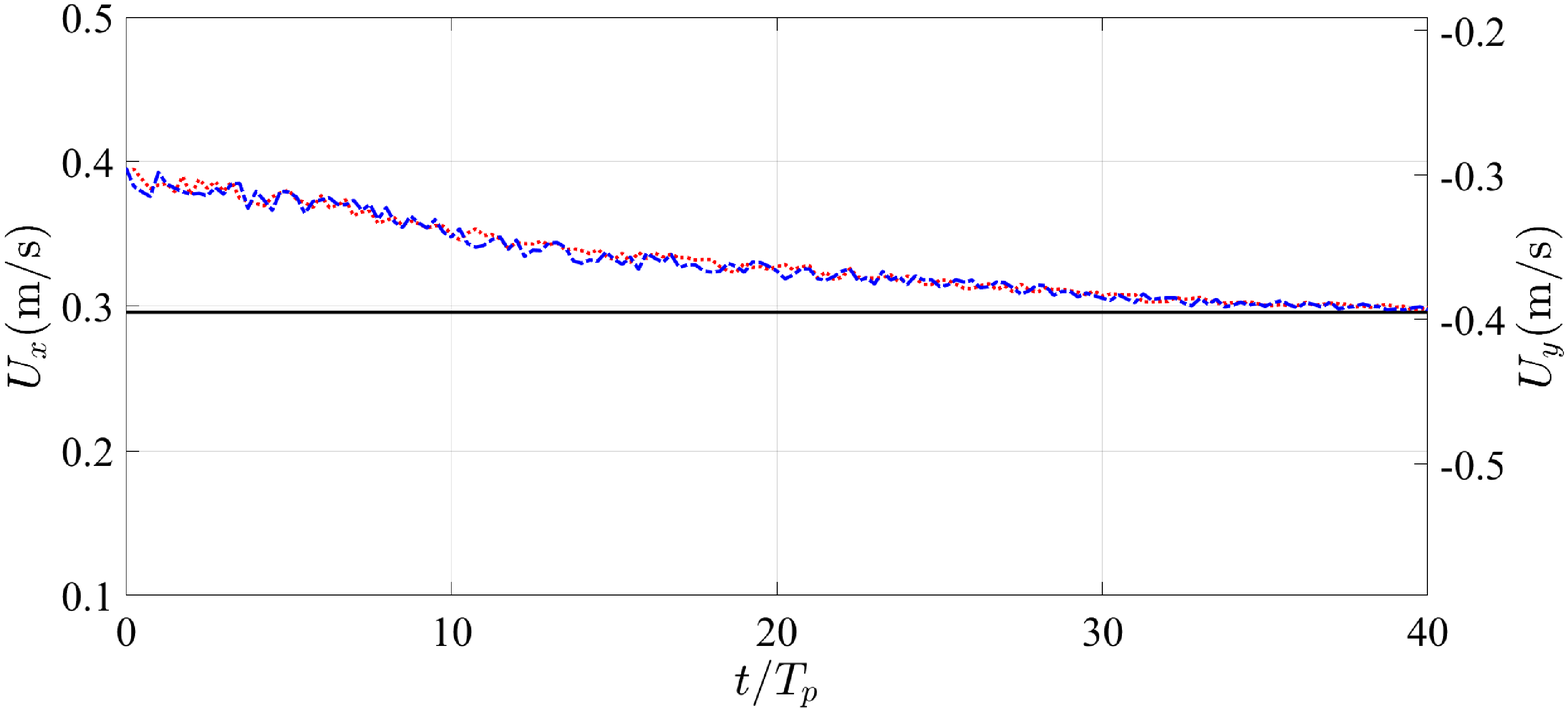}
         \caption{}
         \label{fig:ureal2}
     \end{subfigure}
     
          \begin{subfigure}[b]{1\textwidth}
         \centering
         \includegraphics[trim=0cm 0cm 0cm 0cm, clip,scale=0.29]{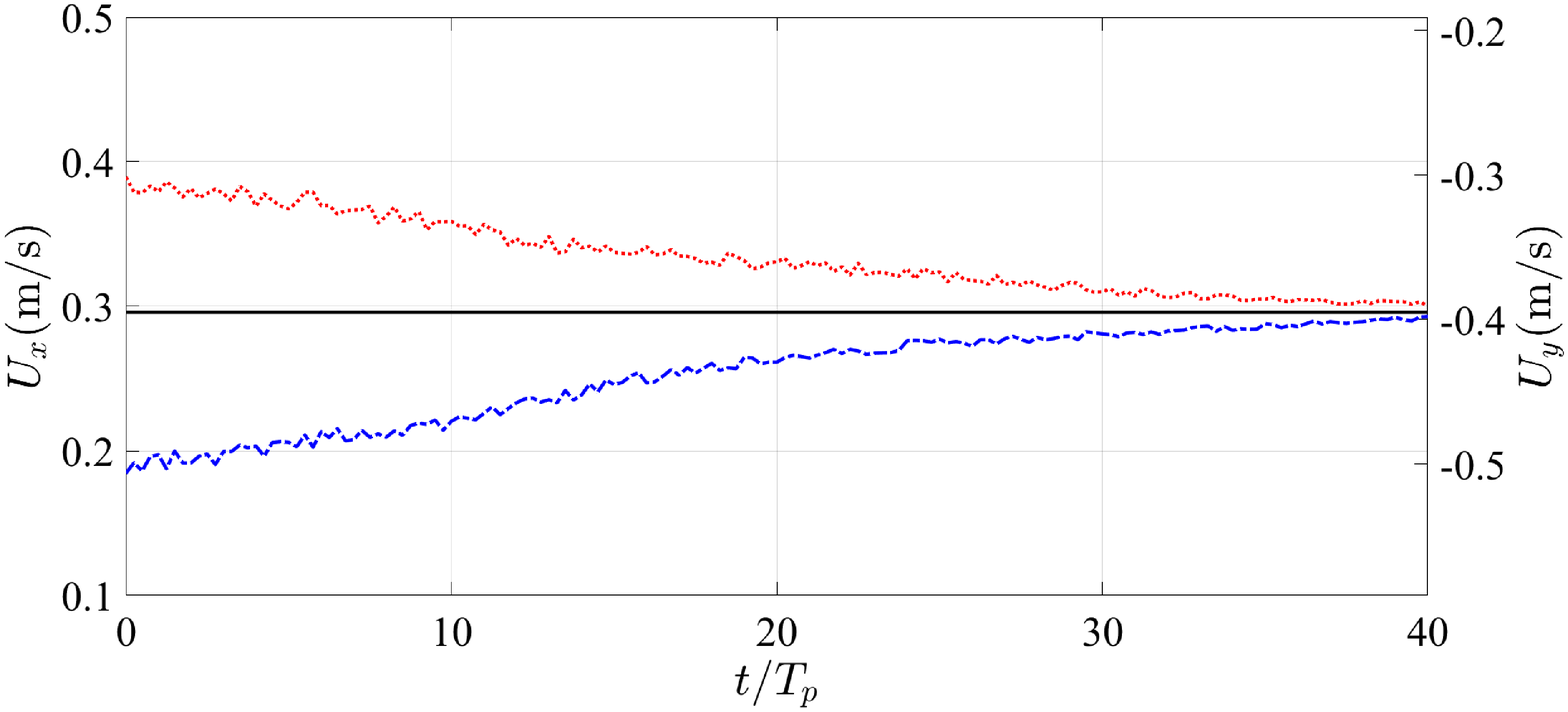}
         \caption{}
         \label{fig:ureal3}
     \end{subfigure}
     
    \begin{subfigure}[b]{1\textwidth}
         \centering
         \includegraphics[trim=0cm 0cm 0cm 0cm, clip,scale=0.29]{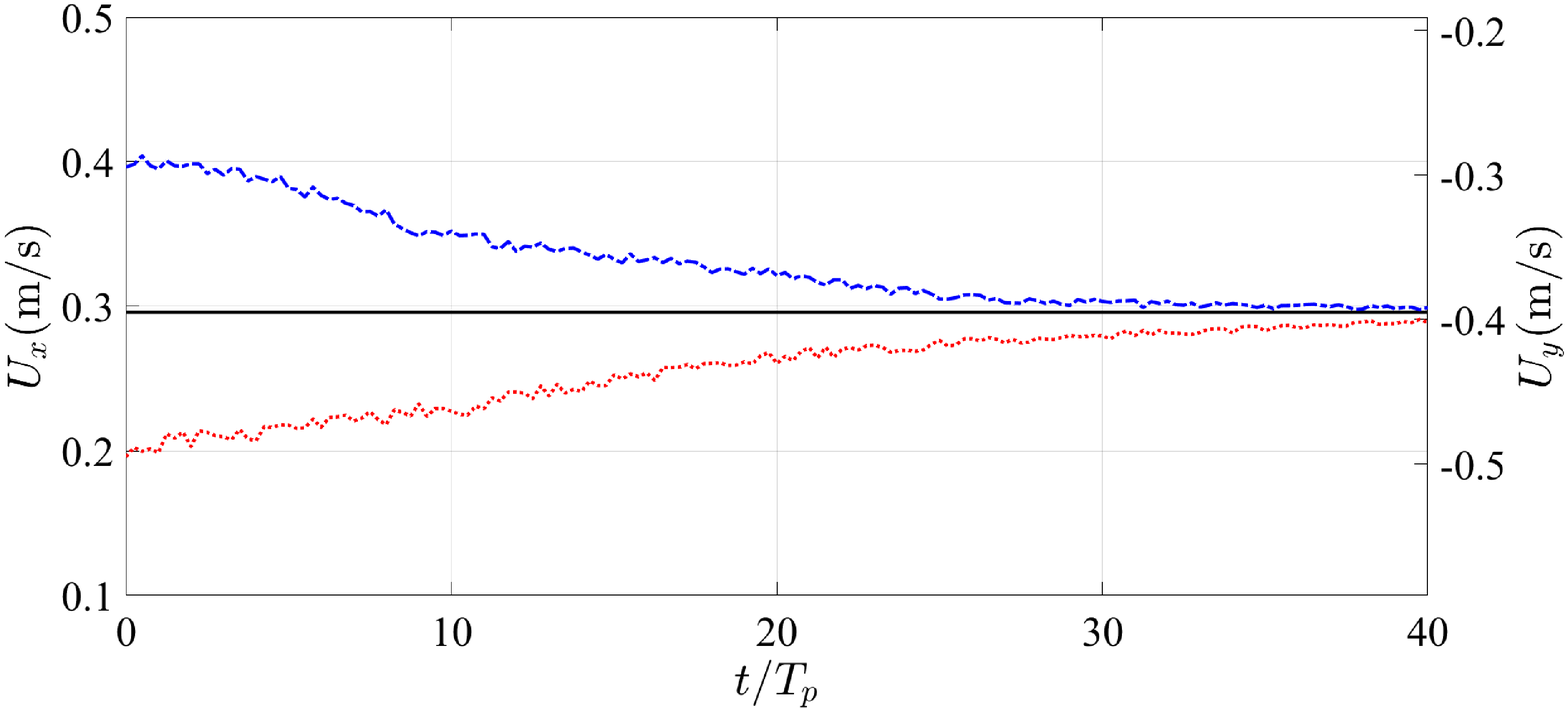}
         \caption{}
         \label{fig:ureal4}
     \end{subfigure}
        \caption{Estimations $U_{x, \text{a}}$({\color{red}\dotL}, left axis) and $U_{y, \text{a}}$ (\hspace{-0.6mm}{\color{blue}\hdashrule[0.5ex]{0.7cm}{0.3mm}{1.3mm 0.5pt 0.4mm 0.5pt}}\hspace{-0.5mm}, right axis) by IEnKF-HOS-C, in comparison with the reference velocity $\boldsymbol{U}_b$ (\L), for four different initial guesses of the current velocity: (a) $\boldsymbol{U}_0=(0.1958,-0.4944)$m/s; (b) $\boldsymbol{U}_0=(0.3958, -0.2944)$m/s; (c) $\boldsymbol{U}_0=(0.3958, -0.4944)$m/s; and (d) $\boldsymbol{U}_0=(0.1958, -0.2944)$m/s.}
        \label{fig:ureal}
\end{figure}

Since the true wave states are inaccessible for this case, we focus on the comparison between the estimated current velocity $\boldsymbol{U}_{\text{a},j}$ and the reference $\boldsymbol{U}_b$ to evaluate the performance of IEnKF-HOS-C. This is also justified from the synthetic cases that an accurate analysis of the wave states is accompanied by a good estimation of the current field. Figure~\ref{fig:ureal} presents the estimation $U_{x,\text{a}}$ and $U_{y,\text{a}}$, in comparison with the reference velocity $\boldsymbol{U}_b$, obtained with different initial guesses (a)~$\sim$~(d). For all the initial guesses, we see that the estimation $\boldsymbol{U}_{\text{a},j}$ converges to the reference $\boldsymbol{U}_b$ at $t_j\approx 40T_p$. The results clearly demonstrate the effectiveness of IEnKF-HOS-C when applied to real radar data, although tests against more sophisticated cases are warranted for future studies (which require better and detailed measurements of the ocean current field together with remote sensing of the surface waves).

\section{Conclusions}
\label{sec:conc}
In this paper, we present a new IEnKF-HOS-C method, which is featured with the capability of simultaneous phase-resolved ocean wave forecast and current estimation. The performance of the IEnKF-HOS-C method is examined using both synthetic data and measurements in the real ocean environment. As indicated by the numerical results, the developed IEnKF-HOS-C method outperforms not only the HOS-C-only method but also the state-of-the-art EnKF-HOS method, in terms of the wave forecast accuracy. In addition, the feasibility of inferring current velocity with this method is extensively demonstrated, by testing it for various forms of the current fields, which are featured with distinct temporal and spatial variations. The developed IEnKF-HOS-C method is intrinsically extensible and can be easily modified to account for other physical or empirical parameters. Finally, if implemented on a graphics processing unit (GPU), this method can be conveniently carried out in offshore environment, which may bring in favorable effects in marine operations.

\section*{Acknowledgement}

G. Wang, J. Zhang, Q. Zhang and Z. Li acknowledge funding from the National Key Research and Development Program of China (2021YFB2601100). G. Wang, J. Zhang, Y. Ma and Z. Li acknowledge funding from the Open Funds of State Key Laboratory of Coastal and Offshore Engineering of China (Project No. LP2101). The authors would like to thank Dr. David Lyzenga in the Department of Naval Architecture and Marine Engineering at the University of Michigan, for providing and interpreting the radar data. 

\section*{Declaration of Interests}
The authors report no conflict of interest.


\bibliographystyle{jfm}
\bibliography{jfm-instructions}

\begin{thebibliography}{32}
\expandafter\ifx\csname natexlab\endcsname\relax\def\natexlab#1{#1}\fi
\def\au#1{#1} \def\ed#1{#1} \def\yr#1{#1}\def\at#1{#1}\def\jt#1{\textit{#1}}
  \def\bt#1{#1}\def\bvol#1{\textbf{#1}} \def\vol#1{#1} \def\pg#1{#1}
  \def\publ#1{#1}\def\arxiv#1{#1}\def\org#1{#1}\def\st#1{\textit{#1}}

\bibitem[Annenkov \& Shrira(2001)]{annenkov2001predictability}
{\sc \au{Annenkov, Sergei~Yu} \& \au{Shrira, Victor~I}} \yr{2001}  \at{On the
  predictability of evolution of surface gravity and gravity--capillary waves}.
   \jt{Physica D: Nonlinear Phenomena}  \bvol{152},  \pg{665--675}.

\bibitem[Aragh \& Nwogu(2008)]{aragh2008variation}
{\sc \au{Aragh, Sina} \& \au{Nwogu, Okey}} \yr{2008}  \at{Variation
  assimilating of synthetic radar data into a pseudo-spectral wave model}.
  \jt{Journal of Coastal Research} ~(52),  \pg{235--244}.

\bibitem[Booij {\em et~al.\/}(1999)Booij, Ris \& Holthuijsen]{booij1999third}
{\sc \au{Booij, NRRC}, \au{Ris, Roeland~C} \& \au{Holthuijsen, Leo~H}}
  \yr{1999}  \at{A third-generation wave model for coastal regions: 1. model
  description and validation}.  \jt{Journal of geophysical research: Oceans}
  \bvol{104}~(C4),  \pg{7649--7666}.

\bibitem[Carrassi {\em et~al.\/}(2018)Carrassi, Bocquet, Bertino \&
  Evensen]{carrassi2018data}
{\sc \au{Carrassi, Alberto}, \au{Bocquet, Marc}, \au{Bertino, Laurent} \&
  \au{Evensen, Geir}} \yr{2018}  \at{Data assimilation in the geosciences: An
  overview of methods, issues, and perspectives}.  \jt{Wiley Interdisciplinary
  Reviews: Climate Change}  \bvol{9}~(5),  \pg{e535}.

\bibitem[Craig \& Sulem(1993)]{craig1993numerical}
{\sc \au{Craig, Walter} \& \au{Sulem, Catherine}} \yr{1993}  \at{Numerical
  simulation of gravity waves}.  \jt{Journal of Computational Physics}
  \bvol{108}~(1),  \pg{73--83}.

\bibitem[Dommermuth \& Yue(1987)]{dommermuth1987high}
{\sc \au{Dommermuth, Douglas~G} \& \au{Yue, Dick~KP}} \yr{1987}  \at{A
  high-order spectral method for the study of nonlinear gravity waves}.
  \jt{Journal of Fluid Mechanics}  \bvol{184},  \pg{267--288}.

\bibitem[Ducrozet {\em et~al.\/}(2021)Ducrozet, Abdolahpour, Nelli \&
  Toffoli]{ducrozet2021predicting}
{\sc \au{Ducrozet, Guillaume}, \au{Abdolahpour, Maryam}, \au{Nelli, Filippo} \&
  \au{Toffoli, Alessandro}} \yr{2021}  \at{Predicting the occurrence of rogue
  waves in the presence of opposing currents with a high-order spectral
  method}.  \jt{Physical Review Fluids}  \bvol{6}~(6),  \pg{064803}.

\bibitem[Evensen(2003)]{evensen2003ensemble}
{\sc \au{Evensen, Geir}} \yr{2003}  \at{The ensemble kalman filter: Theoretical
  formulation and practical implementation}.  \jt{Ocean dynamics}
  \bvol{53}~(4),  \pg{343--367}.

\bibitem[Evensen(2009)]{evensen2009data}
{\sc \au{Evensen, Geir}} \yr{2009} {\em Data assimilation: the ensemble Kalman
  filter\/}.  \publ{Springer Science \& Business Media}.

\bibitem[Fujimoto \& Waseda(2020)]{fujimoto2020ensemble}
{\sc \au{Fujimoto, Wataru} \& \au{Waseda, Takuji}} \yr{2020}
  \at{Ensemble-based variational method for nonlinear inversion of surface
  gravity waves}.  \jt{Journal of Atmospheric and Oceanic Technology}
  \bvol{37}~(1),  \pg{17--31}.

\bibitem[Iglesias {\em et~al.\/}(2013)Iglesias, Law \&
  Stuart]{iglesias2013ensemble}
{\sc \au{Iglesias, Marco~A}, \au{Law, Kody~JH} \& \au{Stuart, Andrew~M}}
  \yr{2013}  \at{Ensemble kalman methods for inverse problems}.  \jt{Inverse
  Problems}  \bvol{29}~(4),  \pg{045001}.

\bibitem[Janssen(2008)]{janssen2008progress}
{\sc \au{Janssen, Peter~AEM}} \yr{2008}  \at{Progress in ocean wave
  forecasting}.  \jt{Journal of Computational Physics}  \bvol{227}~(7),
  \pg{3572--3594}.

\bibitem[Lyzenga {\em et~al.\/}(2015)Lyzenga, Nwogu, Beck, O'Brien, Johnson,
  de~Paolo \& Terrill]{lyzenga2015real}
{\sc \au{Lyzenga, David~R}, \au{Nwogu, Okey~G}, \au{Beck, Robert~F},
  \au{O'Brien, Andrew}, \au{Johnson, Joel}, \au{de~Paolo, Tony} \& \au{Terrill,
  Eric}} \yr{2015} Real-time estimation of ocean wave fields from marine radar
  data.  \bt{In {\em 2015 IEEE International Geoscience and Remote Sensing
  Symposium (IGARSS)\/}},  \pg{pp. 3622--3625}. IEEE.

\bibitem[Ma {\em et~al.\/}(2018)Ma, Sclavounos, Cross-Whiter \&
  Arora]{ma2018wave}
{\sc \au{Ma, Yu}, \au{Sclavounos, Paul~D}, \au{Cross-Whiter, John} \&
  \au{Arora, Dhiraj}} \yr{2018}  \at{Wave forecast and its application to the
  optimal control of offshore floating wind turbine for load mitigation}.
  \jt{Renewable Energy}  \bvol{128},  \pg{163--176}.

\bibitem[Mohaghegh {\em et~al.\/}(2021)Mohaghegh, Murthy \&
  Alam]{mohaghegh2021rapid}
{\sc \au{Mohaghegh, Fazlolah}, \au{Murthy, Jayathi} \& \au{Alam,
  Mohammad-Reza}} \yr{2021}  \at{Rapid phase-resolved prediction of nonlinear
  dispersive waves using machine learning}.  \jt{Applied Ocean Research}
  \bvol{117},  \pg{102920}.

\bibitem[Notton \& Voyant(2018)]{notton2018forecasting}
{\sc \au{Notton, Gilles} \& \au{Voyant, Cyril}} \yr{2018}  \at{Forecasting of
  intermittent solar energy resource}.  \bt{In {\em Advances in Renewable
  Energies and Power Technologies\/}},  \pg{pp. 77--114}.  \publ{Elsevier}.

\bibitem[Nwogu \& Lyzenga(2010)]{nwogu2010surface}
{\sc \au{Nwogu, Okey~G} \& \au{Lyzenga, David~R}} \yr{2010}
  \at{Surface-wavefield estimation from coherent marine radars}.  \jt{IEEE
  Geoscience and Remote Sensing Letters}  \bvol{7}~(4),  \pg{631--635}.

\bibitem[Onorato {\em et~al.\/}(2011)Onorato, Proment \&
  Toffoli]{onorato2011triggering}
{\sc \au{Onorato, Miguel}, \au{Proment, Davide} \& \au{Toffoli, Alessandro}}
  \yr{2011}  \at{Triggering rogue waves in opposing currents}.  \jt{Physical
  review letters}  \bvol{107}~(18),  \pg{184502}.

\bibitem[Pan(2020)]{pan2020model}
{\sc \au{Pan, Yulin}} \yr{2020}  \at{On the model formulations for the
  interaction of nonlinear waves and current}.  \jt{Wave Motion}  \bvol{96},
  \pg{102587}.

\bibitem[Qi {\em et~al.\/}(2018)Qi, Wu, Liu, Kim \& Yue]{qi2018nonlinear}
{\sc \au{Qi, Yusheng}, \au{Wu, Guangyu}, \au{Liu, Yuming}, \au{Kim, Moo-Hyun}
  \& \au{Yue, Dick~KP}} \yr{2018}  \at{Nonlinear phase-resolved reconstruction
  of irregular water waves}.  \jt{Journal of Fluid Mechanics}  \bvol{838},
  \pg{544}.

\bibitem[Santitissadeekorn \& Jones(2015)]{santitissadeekorn2015two}
{\sc \au{Santitissadeekorn, Naratip} \& \au{Jones, Christopher}} \yr{2015}
  \at{Two-stage filtering for joint state-parameter estimation}.  \jt{Monthly
  Weather Review}  \bvol{143}~(6),  \pg{2028--2042}.

\bibitem[Stuhlmeier \& Stiassnie(2021)]{stuhlmeier2021deterministic}
{\sc \au{Stuhlmeier, Raphael} \& \au{Stiassnie, Michael}} \yr{2021}
  \at{Deterministic wave forecasting with the zakharov equation}.  \jt{Journal
  of Fluid Mechanics}  \bvol{913}.

\bibitem[Tolman {\em et~al.\/}(2009)]{tolman2009user}
{\sc \au{Tolman, Hendrik~L} \& \au{others}} \yr{2009}  \at{User manual and
  system documentation of wavewatch iii tm version 3.14}.  \jt{Technical note,
  MMAB Contribution}  \bvol{276},  \pg{220}.

\bibitem[Wang \& Pan(2021)]{wang2021phase}
{\sc \au{Wang, Guangyao} \& \au{Pan, Yulin}} \yr{2021}  \at{Phase-resolved
  ocean wave forecast with ensemble-based data assimilation}.  \jt{Journal of
  Fluid Mechanics}  \bvol{918}.

\bibitem[Wang {\em et~al.\/}(2018)Wang, Ma \& Yan]{wang2018fully}
{\sc \au{Wang, Jinghua}, \au{Ma, Qingwei} \& \au{Yan, Shiqiang}} \yr{2018}
  \at{A fully nonlinear numerical method for modeling wave--current
  interactions}.  \jt{Journal of Computational Physics}  \bvol{369},
  \pg{173--190}.

\bibitem[Wang \& Xiao(2016)]{wang2016data}
{\sc \au{Wang, Jian-Xun} \& \au{Xiao, Heng}} \yr{2016}  \at{Data-driven cfd
  modeling of turbulent flows through complex structures}.  \jt{International
  Journal of Heat and Fluid Flow}  \bvol{62},  \pg{138--149}.

\bibitem[West {\em et~al.\/}(1987)West, Brueckner, Janda, Milder \&
  Milton]{west1987new}
{\sc \au{West, Bruce~J}, \au{Brueckner, Keith~A}, \au{Janda, Ralph~S},
  \au{Milder, D~Michael} \& \au{Milton, Robert~L}} \yr{1987}  \at{A new
  numerical method for surface hydrodynamics}.  \jt{Journal of Geophysical
  Research: Oceans}  \bvol{92}~(C11),  \pg{11803--11824}.

\bibitem[Wu {\em et~al.\/}(2022)Wu, Hao \& Shen]{wu2022improved}
{\sc \au{Wu, Jie}, \au{Hao, Xuanting} \& \au{Shen, Lian}} \yr{2022}  \at{An
  improved adjoint-based ocean wave reconstruction and prediction method}.
  \jt{Flow}  \bvol{2}.

\bibitem[Wu {\em et~al.\/}(2019)Wu, Wang \& Shadden]{wu2019improving}
{\sc \au{Wu, Jiacheng}, \au{Wang, Jian-Xun} \& \au{Shadden, Shawn~C}} \yr{2019}
   \at{Improving the convergence of the iterative ensemble kalman filter by
  resampling}.  \jt{arXiv preprint arXiv:1910.04247} .

\bibitem[Xiao \& Pan(2021)]{xiao2021time}
{\sc \au{Xiao, YMH} \& \au{Pan, Y}} \yr{2021} Time-optimal path planning in an
  evolving ocean wave field based on reachability theory.  \bt{In {\em 2021
  60th IEEE Conference on Decision and Control (CDC)\/}},  \pg{pp. 5019--5026}.
  IEEE.

\bibitem[Xu \& Guyenne(2009)]{xu2009numerical}
{\sc \au{Xu, Liwei} \& \au{Guyenne, Philippe}} \yr{2009}  \at{Numerical
  simulation of three-dimensional nonlinear water waves}.  \jt{Journal of
  Computational Physics}  \bvol{228}~(22),  \pg{8446--8466}.

\bibitem[Yoon {\em et~al.\/}(2015)Yoon, Kim \& Choi]{yoon2015explicit}
{\sc \au{Yoon, Seongjin}, \au{Kim, Jinwhan} \& \au{Choi, Wooyoung}} \yr{2015}
  \at{An explicit data assimilation scheme for a nonlinear wave prediction
  model based on a pseudo-spectral method}.  \jt{IEEE Journal of oceanic
  engineering}  \bvol{41}~(1),  \pg{112--122}.

\end{thebibliography}

\end{document}